\documentclass[aoas,MSNbibl,nameyear,seceqn,dvips]{arximspdf}
\usepackage{graphicx}
\doi{10.1214/14-AOAS745} %kopijuoti is PTS
\volume{8}
\issue{3}
\pubyear{2014}
\firstpage{1640}
\lastpage{1670}
\docsubty{FLA}

\makeatletter
\def\mid{|}
\renewcommand{\mathrm}{\mathit}
\def\smallsetminus{\setminus}

\def\Pi{\prod}
\def\cal{\mathcal}
\newproclaim{remark}{Remark}[section]
\newproclaim{example}{Example}[section]
\newproclaim{assumption}{Assumption}[section]

\def\i{\item}
\def\cald{{\cal D}}
\makeatother

\begin{document}
\begin{frontmatter}

\title{Reducing estimation bias in adaptively changing monitoring
networks with preferential site selection}
\runtitle{Adjusting estimates for preferential sampling}

\begin{aug}
\author[A]{\fnms{James V.}~\snm{Zidek}\corref{}\thanksref{m1,mm1}\ead[label=e1]{jim@stat.ubc.ca}},
\author[B]{\fnms{Gavin}~\snm{Shaddick}\thanksref{m2}\ead[label=e2]{masgs@bath.ac.uk}}
\and
\author[A]{\fnms{Carolyn G.}~\snm{Taylor}\ead[label=e3]{cgtaylor@stat.ubc.ca}\thanksref{m1,mm1}}
\affiliation{University of British Columbia\thanksmark{m1} and
University of Bath\thanksmark{m2}}
\thankstext{mm1}{Supported in part by the Natural Science and
Engineering Research Council of Canada.}
\runauthor{J.~V. Zidek, G. Shaddick and C.~G. Taylor}

\address[A]{J.~V. Zidek\\
C.~G. Taylor\\
Department of Statistics\\
University of British Columbia\\
2207 Main Mall\\
Vancouver, British Columbia V6T 1Z4\\
Canada\\
\printead{e1}\\
\phantom{E-mail:\ }\printead*{e3}}

\address[B]{G. Shaddick\\
Department of Mathematical Sciences\\
University of Bath\\
Bath, BA2 7AY\\
United Kingdom\\
\printead{e2}}

%University of British Columbia\\
%2207 Main Mall
%Vancouver, BC\\
%V6T 1Z4, Canada\\

\end{aug}

% HISTORY:
\received{\smonth{9} \syear{2012}}
\revised{\smonth{4} \syear{2014}}

% ABSTRACT
%
\begin{abstract}
This paper explores the topic of preferential sampling, specifically
situations where
monitoring sites in environmental networks are preferentially located
by the designers. This means the data arising from such networks may
not accurately characterize the spatio-temporal field they intend to
monitor. Approaches that have been developed to mitigate the effects of
preferential sampling in various contexts are reviewed and, building on
these approaches, a general framework for dealing with the effects of
preferential sampling in environmental monitoring is proposed.
Strategies for implementation are proposed, leading to a method for
improving the accuracy of official statistics used to report trends and
inform regulatory policy. An essential feature of the method is its
capacity to learn the preferential selection process over time and
hence to reduce bias in these statistics. Simulation studies suggest
dramatic reductions in bias are possible. A case study demonstrates use
of the method in assessing the levels of air pollution due to black
smoke in the UK over an extended period (1970--1996). In particular,
dramatic reductions in the estimates of the number of sites out of
compliance are observed.
\end{abstract}

% KEYWORDS
% Pirmas kwd is didziosios raides
%
\end{frontmatter}

%s1 #&#
\section{Introduction}\label{sect:intro}\label{sect:introduction}
This paper
addresses the location of monitoring sites within environmental
monitoring networks. In many cases, sampling locations may be dependent
on the responses themselves or parameters of the underlying
environmental process, leading to what \citet{diggle2010geostatistical}
(hereafter referred to as D10) refers to as ``preferential sampling.''
Since measurements from these
sites may be critical for informing policy, assessing adherence to
standards or for health analyses, the potential effects of such
sampling may be of concern. For example,
urban air pollution monitoring sites provide information
that may be used to detect noncompliance with air quality
standards [\citet{ozone05}]. The designer may then
locate the sites where air pollution levels
are believed to be the highest, although
reaching that goal presents its own challenges
as shown in \citet{chang2007designing}.
Reaching this goal would mean
the measured concentrations would
overestimate the levels of the pollutant in that urban area. That
could render these data unsuitable for other
purposes, for example, in epidemiological studies estimating risks to health.

The focus here is on the calculation of official statistics, where both
simplicity and transparency
are important. Such statistics traditionally estimate population
averages, totals and proportions. Their widespread use and importance
means that unbiased estimates are essential and that aggregates of such
statistics will also be unbiased. In this paper we propose an approach
based on the Horvitz--Thompson (HT) estimator, which has played a
key role in the theory of survey sampling, to producing unbiased
estimates of such statistics. The approach builds on the idea of
response biased sampling surveyed in
\citet{scott11} (hereafter S11),
which extends the work in
\citet{lawless1999semiparametric}.

The work was motivated by changes in a large scale air pollution
monitoring network in the UK. The network measured Black Smoke (BS), a
measure of fine particulate matter, and was in operation for more than
fifty years. The very high annual concentrations
seen in the early part of its operation led to successful mitigation
measures and a dramatic decline in those levels. As a result, the need
to monitor BS decreased
and the number of sites was reduced. At its peak in the 1960s, there
were over 1000 sites, but of these only 35 were still operational in
the mid-nineties. The sites that were removed from the network had
lower concentrations than those that remained and the (small number of)
sites that were added to the network over this time had higher levels
[\citet{shaddick14}]. There is therefore clear evidence of preferential
sampling over time and, thus, the decline in reported annual levels of
BS was systematically
underestimated. This in turn means that exceedances of statutory limits
may be overestimated and estimates of health effects of BS may be biased.

A primary aim of this paper is to develop a methodology to adjust
annual averages of BS and the proportions of regions
in noncompliance with criteria imposed as part of the mitigation
strategies for preferential sampling. It is more generally
about approaches to correcting for the deleterious effects of
preferential sampling on population parameter estimates. In order to
address these issues, a new theory is developed and assessed.

The remainder of the paper is organized as follows. Section~\ref
{sect:background} provides a background to previous approaches to
preferential sampling in both nonspatial and spatial settings with
consideration of issues commonly encountered in environmental modeling.
Section~\ref{sect:superpopmodel} provides a superpopulation framework for
building a unified approach to dealing with the effects of preferential
selection. This is followed in Section~\ref{sect:implementation} by strategies
for implementation based on the HT approach.
Section~\ref{sect:simulations} demonstrates the use of the proposed
methods using simulation studies which are then applied in Section~\ref
{sect:blacksmoke} to the
case study of changes in the UK BS monitoring network from 1970 to
1996. In these examples,
we show that correcting for bias can substantially
reduce estimates of the number of sites, monitored and unmonitored,
that are out of compliance with regulatory standards. Section~\ref
{sect:conclusions} discusses our findings and provides some suggestions
for alternative approaches to mitigating the effects of preferential
sampling in a number of settings.

%s2 #&#
\section{Background}\label{sect:background}

In this section we review previous approaches that have been proposed
to mitigate the effects of preferential sampling. We start with methods
that were originally proposed in a nonspatial context but which we
contend can contribute to the development of the field and which we
relate to a spatial setting. This is followed by a review of methods
which have recently been proposed in the field of spatial modeling. The
methodology that we propose here, which is unique in this setting in
its use of temporal changes to infer the levels of preferential
sampling, draws on aspects of many of these approaches and this is
discussed at the end of the section.

We note the distinction between the design-based and model-based
approaches to
spatial design and inference as described by \citet{cicchitelli12}, who
argue that the former is
appropriate when inference relates to global
quantities, such as means or totals, and
the latter when
``constructing a map,'' that is, when performing
spatial prediction or interpolation. The approaches that have been
used to develop monitoring
networks may be classified as follows: (1) unknown; (2) a combination of
networks each developed by an unknown approach [\citet{zidek2000designing}];
(3) unknown in detail but subject to guidelines;
(4) design-based (multi-stage surveys) [\citet{usepa2009national}]; (5)
model-based [\citet{schumacher1993using}]; (6) model-based redesign of
networks designed by unknown approaches \citet{ainslieapplication2009}. The
approaches are diverse or unknown, but statistical models can be
created to model the results and explore the bias in the outcomes.

Throughout the section, we consider a response, $\mathrm{Y}$, and a set of
covariates or explanatory variables, $\mathrm{X}$, which in a spatial setting
would be indexed by their spatial
locations $u_j$. Often interest is in estimating the association
between $\mathrm{Y}$ and $\mathrm{X}$, represented by $\beta$ where $f(\mathrm{Y}) \sim( \beta_0 +
\beta\mathrm{X})$.

\subsection*{Response-biased regression modeling}
Here we consider the possibility for bias in estimating relationships
between a response $\mathrm{Y}$ and a set of covariates or explanatory variables,
$\mathrm{X}$, when the sample of data may be subject to preferential sampling.
This is referred to as modeling ``with response-biased samples'' by
\citet{scott11} (hereafter S11), who extend the work of
\citet{lawless1999semiparametric}. This
has origins in case--control observational studies where the
response $\mathrm{Y}$ is a ``case'' or a ``control'' with
$\mathrm{X}$ being observed (and thus available for analysis) for a sample from
the population
of cases and controls. Models here
assume a finite population of
possible sample items.

Two approaches for inference are suggested in
S11 and we refer to them as follows:

\begin{enumerate}[CML:]
\item[HT:] the Horvitz--Thompson (HT) approach [\citeauthor{horvitz1952}\break (\citeyear{horvitz1952})]:

\begin{enumerate}[$-$]
\item[$-$] uses
estimating equations
designed to construct
design-unbiased estimators when finite population
elements have unequal probabilities of being selected
in a design-based analysis.
\end{enumerate}
\item[CML:] the conditional maximum likelihood (CML)
approach:

\begin{enumerate}
\item[$-$] based on the profile likelihood found by maximizing the joint
distribution
over all possible marginal distributions.
\end{enumerate}
\end{enumerate}
We now describe each of these methods in turn, starting with the HT,
and consider how they might adapt to be used in a spatial setting.

Define
$ \mathrm{R} $ as the sampled site indicator
such that $\mathrm{R}_u $ is $ 1 $ or $ 0 $ according
to whether site $\mathrm{u}$ is selected into the sample or not. Let
\[
\pi_\mathrm{u} = \pi(y_\mathrm{u}, x_\mathrm{u} ) = P\{
R_\mathrm {u} = 1 | y_\mathrm{u,} x_\mathrm{u} \},
\]
the selection probability for site $\mathrm{u}$.
Here we consider the case of
spatial regression where interest will focus on inference on $\beta$, the
coefficient associated with the explanatory variables in the mean
function. Its estimate is found
by using the HT approach and solving the estimating equations
%
%e2.1 #&#
\begin{equation}
\label{eq:HT} \sum_u \frac{ \mathrm{R}_\mathrm{u}}{ \pi_\mathrm{u}}
\frac{\partial\log{ [y_\mathrm{u} | x_\mathrm{u}, \beta] } }{
\partial
\beta} = 0,
\end{equation}
assuming $ \pi_\mathrm{u} > 0, u \in{\cal D}$ are known at
the sampled sites.
When working under the asymptotic paradigm, the selection probabilities
are required to be consistently estimable rather than the requirement
that they be known.

The second, CML, approach is based on the profile likelihood found by
maximizing the joint distribution
over all possible marginal distributions $ [\mathrm{X} ] $ for $X$. Maximizing
the resulting profile yields the estimating equation that characterizes the
CML-approach:
%
%e2.2 #&#
\begin{equation}
\label{eq:PL} \sum_\mathrm{u} \mathrm{R}_\mathrm{u}
\frac{\partial\log{ [y_\mathrm{u} | x_\mathrm{u}, \beta, \mathrm
{R}_\mathrm{u} = 1 ] } }{ \partial\beta} = 0,
\end{equation}
which depends on the $ \{ \pi_\mathrm{u} \} $, each being the
(conditional) probability that $R_u=1$.

In their current form both the HT and CML approaches have limitations
in the setting which we consider here. First, the
assumption that the responses $\mathrm{Y}$ on which
the $\pi_\mathrm{u}$ depend are known seems implausible unlike in
case-controlled studies. Their estimation would require a combination
of the design- and model-based approaches. Second,
the failure
to incorporate spatial correlation is likely to be a serious
limitation. Consistency of the solutions to estimating equations~(\ref
{eq:HT}) and~(\ref{eq:PL}) means
covariances can be estimated by the
``sandwich estimator'' [\citet{rao1998quasi}] and if samples are
sufficiently large, this means the assumption of spatial independence
may be avoided. However, this may be of limited use since many applications
such as spatial prediction rely on the estimated
covariance and estimates of its parameters.

\citet{cicchitelli12} propose a design-based approach which
specifically addresses the issue of spatial dependence. This is based
on the premise that spatial dependence is a result of unmeasured or
unrecognized covariates, a view articulated in D10's discussion
section. Sites in close proximity to one another will be influenced by
local environmental factors
that reflect a spatial pattern and, if not included in a model, these
factors will induce a spatial pattern. The (measured) covariate matrix, $\mathrm{X}$,
is augmented with a matrix of quasi-covariates, $\mathrm{Z}$. For each location,
$u_j$, the quasi-covariate values in $\mathrm{Z}$, $\mathrm{z}_{u_j} = (\mathrm
{z}_{1u_j},\ldots,\mathrm{z}_{Ku_j})$, are modeled using thin-plate
splines with $\mathrm{z}_{u_j} = \tilde{\mathrm{z}}_{u_j} \Xi^{-1/2}$,
$\tilde{\mathrm{z}}_{ku_j} =
(\Vert \mathrm{u}_j - \kappa_k \Vert )^2 \log{(\Vert \mathrm{u}_j - \kappa_k \Vert )},
k= 1,\ldots, K $,
where the $\{\kappa_k \}$ represent representative sets of spatial
locations (knots) such that $\Xi$ is nonsingular. The response is then
modeled as a combination of the measured and unmeasured covariates,
\[
E_\xi(Y_{u_j}) = \beta_0 +\beta_1
x_{1u_j}+ \cdots+ \gamma_1 \mathrm{z}_{1u_j} +
\cdots+ \gamma_K \mathrm{z}_{Ku_j},
\]
where $E_\xi$ denotes the expected value with respect to the model
$\xi
$ (in terms of the material introduced in Section~\ref
{sect:superpopmodel}, this is the superpopulation model's distribution
for $Y$ from which the finite population is drawn).
They then invoke the design-based approach and use
HT estimators to fit a regression model for
Y on $(\mathrm{X},\mathrm{Z})$. In line with the HT approach, they
do not explicitly model spatial structure in their model-assisted,
design-based
approach, arguing that spatial pattern is provided
by the mean function with the augmented covariates.
Their position is supported by the
well-known duality between
first order and second order features of geostatistical models;
misspecification of the first order mean function will always lead to
bias in the second order variogram.

\subsection*{Spatial prediction}
We now consider the model-based approach as used in [D10: \citet
{diggle2010geostatistical}, \citet{gelfand2012effect}, \citet
{pati2011bayesian}], all of which specifically consider spatial
modelling. D10, which includes
a bibliography of earlier work on preferential sampling unrelated
to this paper, characterizes the effect
of preferential sampling within a model-based geostatistical framework
in which
site locations are informative for inference.
D10 assumes a latent, unobservable Gaussian field $\mathrm{S}$ over a
geographical continuum (domain) ${\cal D}$. The sites $\mathrm{u}$ are
selected at random in accordance with an inhomogeneous
Poisson spatial process with intensity function
$\lambda_u = \exp{\{\alpha+ \beta\mathrm{S}_u\}, u\in{\cal D}}$.
The measurable response $\mathrm{Y}$ is also modeled as dependent on
$\mathrm{S}$. Both the response measurements, $y^{\scriptsize\mbox{obs}}$, and selected sites $\mathrm{U}$
yield information about the underlying model parameters, including both
those in the spatial mean as well as the spatial covariance matrix for $\mathrm{S}$.
A marginal likelihood function is obtained by marginalizing out
$\mathrm{S}$ although $\mathrm{S}$, $\mathrm{U}$ and $\mathrm{Y}$ are correlated in their joint spatial distribution.
A key to the success of the model in D10 is
knowledge of process and data, both present and future, in order to
characterize the sampling selection process, an assumption questioned
by \citet{dawid:2010}. Given that knowledge, the effects of preferential
sampling on variogram estimates,
spatial predictions and associated biases can be assessed. \citet
{pati2011bayesian} extend
that approach, again relying on latent variables in a point process approach.
\citet{gelfand2012effect} suggest an alternative approach based on
knowledge of the underlying process being monitored and the factors
that drive that process.

\subsection*{The temporal dimension} The approaches described
above do not include changes to networks due to preferential
sampling over time, a key feature of spatial sampling addressed in this paper.
Examples of this include the redesign of Vancouver's
air quality monitoring network [\citet{ainslieapplication2009}] and the
case study in Section~\ref{sect:blacksmoke} which considers changes in
a long-term air pollution monitoring network in the UK. \citet
{le2006statistical}
demonstrate the use of such adaption as the
network for monitoring ground level ozone concentrations that has
been steadily augmented over the last few decades as the adverse health
effects of ozone have been recognized. In this example, it
seems plausible that the addition of sites has been performed preferentially
to ensure high levels of ozone are detected.
The case study in this paper considers
the monitoring of black smoke (BS) for which the number
of monitoring sites declined from ca. 1000 in the early 1960s to
ca. 100 at the turn of the last century as
levels of BS declined due to improvements in the management of
air quality. \citet{shaddick14} demonstrate that the reduction in the
network was done preferentially by removing
sampling sites with generally lower concentrations
relative to the decline in overall levels of BS, leading to
overestimates of BS concentrations (as can be seen in Section~\ref
{sect:blacksmoke}). We propose that by considering the stochastic
process of
selection over time, it is possible to model the extent of preferential
sampling and estimate its deleterious
effects on published estimates of environmental fields and the
effects on items of inferential interest. This emphasis leads us to
build on the HT approach and the use
of estimating equations following the celebrated work in
this area by David Binder [\citet{binder1983variances}].

Unlike the earlier methods previously described, this approach
incorporates time in the model. We draw on the previous methods and
specifically those based on the H-T estimator, and assume that
the sampling probabilities are uncertain and model the selection
process as stochastic, depending sequentially on the
responses over time. The result is an approach that is a combination of
the design- and model-based approaches. The $\{\pi\}$ are then learned
over time as the results of monitoring accumulate.

Populations of potential
sampling sites can often be taken, as in this
paper, to be a finite set of possible locations.
In that sense, this approach diverges from D10 and its successors in
the model-based category
by assuming a finite population of $\mathrm{N}$ sampling (i.e., monitoring) sites,
$\mathrm{u}_j, j=1,\ldots, {\mathrm N} $.
%%and also confront the computational problems
%%when N is large.
Note that $\mathrm{u}_j$ could represent
just the label $j$ or, more commonly, the geographic
coordinates of the site, depending on the context.
This is not to say that the assumption of a continuous domain
for site selection is unreasonable in all cases, but
practical and administrative considerations will
often restrict $\mathrm{S}$ to be a finite-dimensional vector-valued process
over a discrete
domain ${\cal D}$.
In D10 $\mathrm{S}$ is also discretized, but this is in order to approximate
the marginal likelihood and is done by
replacing the
continuous ${\cal D}$ by a fairly dense lattice.
Sampling points then have to be mapped onto
their nearest lattice point neighbors.

As with the successors to D10 cited above, we
allow covariates to be incorporated.
Site selection
may well depend on them, for example, the distance of a site from a
major roadway or an ``urban-rural'' classification (as used in the
case study in Section~\ref{sect:blacksmoke}). In addition,
interest may well lie in the significance of the
effects of such covariates or design variables on the
measured responses.

%s3 #&#
\section{A general framework}\label{sect:superpopmodel}

Section~\ref{sect:background} presents a number of paradigms
in which to study the issue of preferential spatial sampling within
the design- or model-based frameworks. The latter includes
the Bayesian approach although it was not explicitly mentioned.
The concept of a superpopulation [\citet{sarndal03}] provides a
framework for
unifying paradigms for inference and it is within this framework that
we develop the proposed approach.
We combine design- and model-based approaches, allowing us to estimate
the unknown
selection probabilities needed
for the HT estimator which are used to compensate for the selection
bias introduced in sampling from the finite population of interest.

We suppose a discrete
geographical domain ${\cal D}$ contains point referenced
sites $\mathrm{u}_ j, j =   1,  \ldots,  \mathrm{N} $.
Let $\mathrm{T}$ denote the present time and although the spatial
locations do not have a natural ordering, it
is convenient to use the vector notation
$ \mathbf{Y}_t\dvtx 1 \times\mathrm{N}$
to represent the sequence of
responses at those sites at times
$ t   =   1,  \ldots,  \mathrm{T} $.
These sites, which need not be on a lattice, represent a
finite population of potential locations
at which to site monitors that repeatedly measure at regular times,
a random space--time field. Further, let $\mathbf{Y}$ denote the
$\mathrm{T}
\times
\mathrm{N}$ matrix
comprised of those row response vectors. Similarly at time $t$, let $
\mathbf{X}_t $
denote a matrix of covariates or explanatory
factors, hereafter referred to as ``covariates''
for simplicity. Then $\mathbf{X}$ denotes the
corresponding three-dimensional covariate array.

We now propose a framework which contains
three major
components in the process of using data from monitoring networks, which
for simplicity in subsequent descriptions we characterize as (i)
Nature, (ii) the Preferential Sampler and (iii) the Statistician:

%heere
%
\begin{longlist}[(iii)]
\item[(i)] Nature governs the process-model, a joint distribution for
$\mathbf{X}$ and $\mathbf{Y}$ that generates realizations
$\mathbf{x}$ and $\mathbf{y}$
over the time period ending
at the present time, $T$. These are regarded
as drawn from an infinite population
of possibilities called the Superpopulation
that is indexed by the sites in ${\cal D}$.
In some contexts this would be the relevant population
and the parameters of the process-model the
objects of inferential interest. However,
here we consider the case, commonly encountered in official statistics,
where there is a finite population of possible site locations and thus
values of $\mathbf{x}$ and $\mathbf{y}$ which we refer
to as the Population.

\item[(ii)] The Preferential Sampler runs the measurement-model, the process
that
chooses sites in ${\cal D}$ at which the associated
elements of $\mathbf{x}$ and $\mathbf{y}$ are observed. The sampling
design, that is, the selection
probabilities for sites to be included in the sample at each time
$t=1,\ldots, T$
may depend on the elements of $\mathbf{x}$ and/or $\mathbf{y}$ for
previous times. The resulting sample
can be interpreted
as from either the Population or Superpopulation
depending on the goal of the monitoring program.

\item[(iii)] The Statistician, working within a design-based framework
and using only the sample and knowledge
of the sites at which the sample was collected, infers the uncertain
selection probabilities
used by the Preferential Sampler and through this
adjusts inferences about the Population
(or Superpopulation as appropriate) to compensate for the
bias induced by preferential sampling.
\end{longlist}

The mechanisms of the preferential sampling process and associated
selection probabilities can be complex and are often not well
understood or may be unknown. They
may be nonanthropogenic or anthropogenic. Examples of the former might include
a natural event, such as a forest fire, making some locations
inaccessible (and thus their selection probabilities zero). Although
responses continue to be generated, they could not be sampled. For the
latter, there may be new guidelines on where sites may be located or
new developments or changes of land ownership may mean sites can no
longer be located at the same places. These uncertain mechanisms
that lead to the sample make the selection probabilities uncertain as
well meaning that they need to be inferred.
The Superpopulation--Population-Sample paradigm provides a framework
where that becomes feasible.

We now formalize the ideas above in a general theory for response generation~($\mathbf{Y}$) and site selection ($\mathbf{R}$ from Section~\ref
{sect:background}),
one that allows flexibility in the choice of modeling, inferential
and selection paradigms.
We first develop a theory based on the
conditional distribution of $\mathbf{Y}$ given $\mathbf{X} = \mathbf{x}$
and model parameters $\theta$. These model parameters
would include all those that characterize the joint distribution and
could include, depending on the context,
regression coefficients $\beta$ [as seen in equation (\ref{eq:HT})] and
autocorrelation and spatial covariance parameters.
Using the notation from D10, we denote
the conditional distribution of $\mathbf{Y}$, which may be
characterized by its probability density or cumulative distribution function,
by
%
%e3.1 #&#
\begin{equation}
\label{eq:superpopdensity} [\mathbf{{y}} | \mathbf{{x} }, \theta].
\end{equation}

If within a specified time period,
all responses and covariates were observed for every
spatial site in the Population,
%%of sites
%%to get a set $ W$,
we could proceed in the usual way to make inferences
about $\theta$.
In particular, given $\mathbf{Y} = \mathbf{y} $
and $\mathbf{X} = \mathbf{x} $,
the conditional likelihood function
would be given by equation (\ref{eq:superpopdensity}).
The Superpopulation's
maximum likelihood estimator (MLE) of $ \theta$, denoted
by $ \hat{\theta} = \theta(\mathbf{y},\mathbf{x} ) $, would
estimate $\theta$, including
temporal as well as spatial
correlation model parameters together with coefficients in the
regression model relating $\mathbf{Y}$ and $\mathbf{X}$.
Alternatively, if only a random sample of sites were selected from
the finite population of sites
and their
associated response-covariate pairs recorded,
%% to yield a set $w$,
an estimate\hspace*{1.8pt} $\hat{\hspace*{-1.8pt}\hat{\theta}}$
of $\hat{\theta}$ could be computed and considered
to be an estimate of $\theta$.

%%%%%%%%%%%%MAJOR CHANGES FOLLOW
%%%%%%%%%%%%%%%%%%%%%%%%%
In contexts where official
statistics are collected and published
or regulatory policy is administered, estimates
for specific times are commonly required.
In such cases the
Population at the present time $T$
consists of just the responses
$\mathbf{Y}_T = \mathbf{y}_T$ generated
by the marginal distribution
%
%e3.2 #&#
\begin{equation}
\label{eq:superpopdensityT} [ \mathbf{Y}_T \mid\mathbf{X}_T =
\mathbf{x}_T, \theta_T],
\end{equation}
where we have assumed $\mathbf{Y}_T$
depends only on the covariates at time $T$.
Hereafter for simplicity
``Population'' will refer generically
to this time dependent population.
%%Th the target might well be the
%%Population rather than the Superpopulation.
%%The $\mathbf{Y} = \mathbf{y} $ and
%%$\mathbf{X} = \mathbf{x}$ pairs
%%are then
%%$w$
%%sub--sampled
%%have been drawn.
%%and inferences drawn about the
%%Population's numerical characteristics.
%%${\cal W}$.
%% rather
%%than $\boldsymbol{\theta}$.
%%Now
%%estimates of the numerical characteristics
%%of the real, finite population
%%of these pairs would be required.
Following a standard approach
in survey sampling theory
[\citet{sarndal03}], we would define
$ \hat{\theta}_T$ to be the matrix of
parameters of the Population, obtained
by maximimizing the marginal likelihood
in equation (\ref{eq:superpopdensityT}),
and take it to be the
object of inference, although
it may also be viewed as
representing $\theta_T$. [For a discussion
of this issue see \citet{pfeffermann93}.]
Therefore, two legitimate objects of inferential
inference present themselves, $\theta_T$ and
$ \hat{\theta}_T$.
In either case, as in D10, we
are concerned with the effects of preferential sampling
on the estimates derived from the sample
of $ \hat{\theta}_T$ and, in turn, $\theta_T$, depending on the
inferential objective.
%%But in either case, we rely
%%on the superpopulation model indexed by the
%%former, which defines the latter, to develop
%%approaches to assess those effects.

To formalize these ideas, we express the Superpopulation
log-likelihood estimating equation for the MLE $\hat{\theta}_T$
as
%
%e3.3 #&#
\begin{equation}
\label{eq:mleeqn} \nabla_{\theta_T} \log{ [ \mathbf{y}_T |
\mathbf{x}_{T}, \theta_T] } = 0.
\end{equation}

%%heere
The measurement-model is more complex
since the process for selecting the sites
at time $T$, on which inference is
to be based at that time, may depend on responses
at previous times. To model the selection
process, we use notation
introduced in Section~\ref{sect:background}. Thus,
we let $ \mathbf{R}$ denote
the $T \times N $ matrix of indicator random variables
whose $t$th  row $\mathbf{R}_t$ consists entirely of zeros except
for ones in the columns corresponding to the sites
selected for inclusion at time $t$.

Let $\mathbf{y}_{\mathbf{r}}$ and $\mathbf{x}_{\mathbf{r}}$ denote the
observed values
of $\mathbf{Y}$ and $\mathbf{X }$ at the design points selected
adaptively over
time. In other words, if $\mathbf{r} = (r_{tj})$, then
\[
\mathbf{y}_{\mathbf{r}} = \{y_{tu_{tj}}\dvtx t, j\mbox{ for which }
r_{tj} = 1\}
\]
and so on.
We model the distribution of $\mathbf{R}_t$
as stochastically dependent only on $\mathbf{Y}_{1:(t-1)}$ and
$\mathbf{X}_{1:(t-1)}$,
where we use the
general notation $a_{r:s} = (a_r,\ldots, a_s)$ for $ r\leq s$ and
the null vector if $ r > s$ for any
object $a$.
That dependence could reflect the effect
of a latent process as in D10.
Then $\pi_{tu} =
P( R_{tu} = 1 | \mathbf{y}_{1:(t-1)}, \mathbf{x}_{1:(t-1)}, \eta
), t=1,\ldots,T $,
where $\eta$ denotes the
matrix of parameters for the
measurement-model.
%The result may then be
%incorporated in the likelihood.
%More explicitly
%assume the design at Time $T$, $\mathbf{R}_T$, depends
%only on previously observed values of $\mathbf{y}_{1:(T-1)}$ and $
Our assumptions imply that
the conditional preferential sampling
distribution of $\mathbf{R}$ is given by
%
%e3.4 #&#
\begin{equation}
\label{eq:rdistribution} [ \mathbf{r} | \mathbf{y},\mathbf{x},\eta] = \Pi_{t = 1}^T
[ \mathbf{r}_t | \mathbf{y}_{\mathbf{r}_{1:(t-1)}}, \mathbf{x}_{\mathbf{r}_{1:(t-1)}},
\mathbf{r}_{1:(t-1)}, \eta].
\end{equation}

%%heere
Combining equations (\ref{eq:superpopdensityT})
and (\ref{eq:rdistribution}) yields
for inference at time $T$
the joint conditional likelihood
%%Section~\ref{sect:background}
%%suggests two approaches to this problem, one like D10 where we
%%model spatial dependence and the other like S11 where we ignore it.
%%\mathbf{X} = \mathbf{x} is
%%, the former approach would lead to the complete
%
%e3.5 #&#
\begin{equation}
\label{eq:jointlikelihood} L( \eta, \theta_T) \doteq[ \mathbf{y}_{T}
\mid \mathbf{x}_{T}, \theta_T ] \Pi_{t = 1}^T
[ \mathbf{r}_t | \mathbf{y}_{\mathbf{r}_{1:(t-1)}}, \mathbf{x}_{\mathbf{{r}}_{1:(t-1)}},
\mathbf{r}_{1:(t-1)}, \eta].
\end{equation}
%
%%This likelihood includes the responses and the preferentially sampled
%%spatial sites indices, both of which may provide information about
%the parameter
%%matrix $\theta$.
We
assume that $\eta$ does not contain
elements of $\theta_T$ and so
the population parameter matrix
remains as that defined in equation (\ref{eq:mleeqn}).
The likelihood here suggests an
approach for fitting the
site selection probabilities once the
sample is obtained:
impute the nonsampled values and
estimate $\eta$ from the likelihood inferred
from the resulting combination of actual and imputed data.
This is the approach used in later sections with a
logistic regression approach.
Note that while the preferential sampling scheme
represented in equation (\ref{eq:jointlikelihood})
is ancillary for the purpose of estimation
of $\theta_T$, within the design-based framework
below for inferring the population parameter
$\theta$, it is very relevant and in fact
it lies at the heart of the HT approach used
there.

Our general framework can be extended
to include a Bayesian approach by
incorporating a prior joint distribution for
$\eta$ and $\theta$.
% In particular, by allowing
%these parameters to have common elements
%we could
%%$ \pi(\eta, \theta) = \pi(\eta, \theta| \varrho)$ on
%%$\eta$ and $\theta$, where $ \varrho$ is
%%a hyperparameter that would need to be specified.
%%The marginal posterior distribution of
%%of Y and R (as in D10) conditional on the covariates
%%would then be
%%\be\label{eq:posterior}
%%\pi(\eta, \theta|
%%\mathbf{y}, \mathbf{r} ) \propto
%%L( \eta, \theta) \pi(\eta, \theta).
%%\ee
%%This approach would embrace that of
%the approach in D10
%%were we to let
%%$ \theta$
%by including a latent random field that
%underlies both the processes that
%generate the measurements and
%the selection of sites.
%%The hyperparameter vector $\varrho$, could
%%be estimated by Type II maximum likelihood to reduce the
%%computational burden a fully Bayesian approach might
%%entail.
This topic is left for future work.

%%%%%%%%%%%%%%%%%%%%%%%%

\textit{General implementation strategies}.
We now describe general strategies for implementing
our general framework
using the HT approach in Section~\ref{sect:background},
leaving Section~\ref{sect:implementation} for
specific techniques.
We demonstrate how the
general framework might work and provide
a
link to what follows in the next section
where the HT approach is developed.
More specifically, we show that the framework
can be used even when we cannot obtain
estimating equations resembling those in (\ref{eq:HT}) when inter-site
dependence is present.
%% feasible to incorporate the preferential
%%sampling mechanism directly into the likelihood.

The population
parameter matrix associated with that of the
superpopulation
process, $ \theta= (\theta_1,\ldots,\theta_T)^\prime$,
commonly has the form
%
%e3.6 #&#
\begin{equation}
\label{eq:simpcase} \hat{\theta}_t = {\mathbf{H}} \biggl\{
N^{-1} \sum_j \bigl({
\mathbf{h}}_1[\mathbf{y}_{tu_j}, \mathbf{x}_{tu_j} ],
\ldots, {\mathbf{h}}_q [{\mathbf y}_{tu_j}, {
\mathbf{x}}_{tu_j}] \bigr) \biggr\}
\end{equation}
for known functions ${\mathbf{H}}$
and $ {\mathbf{h}}_1,\ldots, {\mathbf{h}}_q $.

%%%%%%at the sites $\{ \mathrm{u}_j \}$
Then
if the $\{ \pi_{tu_j}\}$ are known or estimated,
$ \hat{\theta} $ can be estimated by
%
%e3.7 #&#
\begin{equation}
\label{eq:estimator} \hat{\hspace*{-1.8pt}\hat{\theta}}_t = {\mathbf{H}} \biggl\{ \sum
_u \frac{r_{tu} }{N\pi_{tu} } \bigl({\mathbf{h}}_1[{
\mathbf{y}}_{tu}, {\mathbf{x}}_{tu}], \ldots, {
\mathbf{h}}_q [{\mathbf{y}}_{tu}, {\mathbf{x}}_{tu}]
\bigr) \biggr\}.
\end{equation}
Justification for this choice comes from the unbiasedness of
$ \sum_u ( r_{tu} /N\pi_{tu} )\times
{\mathbf{h}}_l [{\mathbf{y}}_{tu}, {\mathbf{x}}_{tu}], l=1,\ldots,q$
as estimates of their corresponding population
averages.

We illustrate this approach in two specific cases: (1) a regression
model where the interest is in the association between $\mathrm{Y}$
and $\mathrm{X}$, a relationship that may evolve over
time; (2) the estimation of
population means
in the presence of spatial dependence.

\textit{Case \textup{1.} Regression of $\mathrm{Y}$ on $\mathrm{X}$}: Here
\[
H({\mathbf{a}}, {\mathbf{b}}) = \pmatrix{
a_1/b_1
\vspace*{2pt}\cr
\vdots
\vspace*{2pt}\cr
a_T/b_T}
\]
for $\mathrm{T}$ dimensional vectors ${\mathbf{a}}$ and ${\mathbf{b}}$
with $h_1 [y_{tu}, x_{tu} ] = y_{tu} x_{tu} $
and $ h_2[y_{tu}, x_{tu} ] = x_{tu}^2$.

\textit{Case \textup{2.} Spatially dependent Gaussian fields}:
Conditional on the mean and covariance structure,
the Superpopulation model is given by a matric-Normal distribution.
That is,
\[
Y \sim N_{T \times N} (\mu, \Omega\otimes\Lambda),
\]
where $\Omega^{T \times T } $ as well as
$ \Lambda^{N \times N}$
are positive definite matrices, $ E(Y_{tu}) \equiv\nu_t, t=1,\ldots,T$
for all $u$, and
$\mu= \nu\otimes\mathbf{1}$ with $\nu= (\nu_1,\ldots, \nu
_T)^\prime$,
$\mathbf{1}^{1 \times N} = (1,\ldots,1)$.
Assuming $\Omega^{-1} = \tau\tau^{\prime}$ is known, the population parameters
$\{\hat{\nu}_t\}$ are easily found to be
%
%e3.8 #&#
\begin{equation}
\label{eq:nuhat} \hat{\nu}_t = \sum_j
\frac{\mathbf{y}_t \tau^j \tau^{j\prime}
\mathbf
{1}^\prime}{\sum_j (\mathbf{1}\tau^j)^2} =\frac{ \mathbf{y}_t
\Omega
^{-1} \mathbf{1}^\prime}{ \mathbf{1} \Omega^{-1} \mathbf{1}^\prime},
\end{equation}
where $\tau= (\tau^1,\ldots, \tau^N)$. This population parameter has
the form given in equation~(\ref{eq:simpcase}) and so can be adjusted
using the HT approach. Note that the profile likelihood for the
covariances $\Omega$ and $\Lambda$ involves a quadratic form in
$\mathbf{y}$, and it yields estimating equations for the population
level MLEs which can be used to estimate them.

This case is more general than it may seem at first glance. The
Gaussian likelihood in this example can be treated as a
quasi-likelihood leading to the GEE approach [\citet
{liang1986longitudinal}] when $\mathbf{Y}$ is not Gaussian. The
so-called working covariance between columns can be taken as
independent in that case with asymptotic justification providing that
the assumption of equal means across sites holds.
Alternatively, we can
model spatial patterns via the covariates, $\mathbf{X}$,
and then assume no spatial structure in the covariance as in
\citet{cicchitelli12} (see Section~\ref{sect:background}).

Another approach to implementing the general framework
is also available when considering regression where
an estimating equation is of the form seen in~(\ref{eq:HT}). For this case we
suppose $\theta_t = (\beta_t,\Psi)$, where $\beta_t$
is the vector of regression parameters. This plays the same
role as it did in equation (\ref{eq:HT}),
that of parametrizing first order effects
embraced by the process
mean function, while $\Psi$ represents the parameters of
the spatial dependence model, that is, the
covariance in the case of a Gaussian field. We denote $\Psi_0^{\scriptsize\mbox{known}}$ as the case when there is
spatial independence,
a diagonal matrix in the case of the Gaussian
field. Note that Section~\ref{sect:background} provides a discussion
of the
assumption of
independence.

From equation (\ref{eq:mleeqn}), we now
get the superpopulation maximum likelihood
estimating equation for inference about the
population at time $T$:
\begin{eqnarray}\label{eq:nospatcorr}
\nonumber
&& \nabla_{\beta_T} \log{ [\mathbf{y}_T |
 \mathbf{x}_{T}, \beta_T,
\Psi_0 ] }
\nonumber
\\
&&\qquad= \sum_u \nabla_{\beta_T} \log{
[y_{Tu} \mid x_{Tu}, \beta_T,
\Psi_0 ] }
\\
&&\qquad= 0.
\nonumber
\end{eqnarray}
In this way the population parameters are defined.

Following S11, we now have two approaches
for finding the sample-based
MLE. The first is provided by the HT approach
in equation (\ref{eq:HT}),
which yields the following estimating equation:
\begin{equation}
\sum_u \frac{ r_{Tu} }{ \pi_{tu} } \nabla_\beta
\log{ [ y_{Tu} \mid x_{ Tu}, \beta_T,
\Psi_0 ] }  = 0.\label{eq:10}
\end{equation}

The CML-approach for
estimating $\beta$ would rely on the following estimating
equation from equation (\ref{eq:PL}), where
\begin{equation} \label{eq:cmle}
%%\nonumber
%% &&
\sum_u
\nabla_{\beta_T} \log{ [ y_{Tu} \mid x_{Tu},
\beta_T,\Psi_0, R_{tu}=1 ] } =
0.
\end{equation}

If responses can be assumed to the temporally independent, then this
simplifies to be of the same form as (\ref{eq:10}).

A third approach to implementing our framework involves generalizing
the maximum likelihood estimating equation
to a general estimating equation
as described in \citet{godambe1986parameters}.
In the case of temporally independent fields, for the superpopulation
case, this becomes
%
%e3.9 #&#
\begin{equation}
\label{eq:esteqn} \sum_u \phi_{Tu}
(y_{Tu}, x_{Tu}, \beta_T) =0
\end{equation}
for some known functions $\{\phi_{Tu}\}$.
Note that under regularity conditions this gradient has a conditional
expectation,
given the superpopulation parameters, equal to zero, a property
referred to an ``unbiasedness.'' In
fact, equation (\ref{eq:esteqn}) can be used to define an estimator
for any choice of kernel $ \phi$ provided it is unbiased. In this way,
\citet{binder94} formulated a general
approach for complex sample surveys based on estimating equations.

%s4 #&#
\section{Implementing the HT approach}\label{sect:implementation}
This paper proposes the HT, or ``inverse probability weighting,'' approach
to compensate for the
bias introduced by preferential sampling.
The approach can be implemented in a variety of ways depending on the context
and inferential paradigm [\citet{kloog2012incorporating}].
A question that arises in
all cases is the role of spatial dependence and
whether or not it needs to be explicitly acknowledged in the chosen model.
The approaches described in Section~\ref{sect:background} suggest the
answer to this
question is not clear cut and the answer will depend on various
factors, including data that are available
and the inferential objectives. Even where spatial dependence should
be incorporated in the superpopulation
model, the specific application may prohibit it. In the example of case
2 (under the assumption that the model is correct), then the
unequally weighted population
mean in equation (\ref{eq:nuhat}) which incorporates
spatial correlation should ideally be estimated. In practice, however,
the equally weighted average
is commonly used (and is the one that is available for the case study
presented in Section~\ref{sect:blacksmoke}). In such cases,
the HT approach must be applied to the available estimate rather than
an idealized one. This section
suggests an approach that covers
all situations that fall within the general framework of
Section~\ref{sect:superpopmodel}, whereas Sections \ref
{sect:simulations} and~\ref{sect:blacksmoke} focus on the equally
weighted average in order to provide a solution to the problem most
likely found in practice.

In practice, the process which selects the monitoring sites is nonrandom
and generally not known. The selection probabilities cannot therefore
be characterized as they are in multi-stage survey sampling, for example.
However,
having a time series of samples of sites from the finite
population of $N$ sites enables us
to model the selection process and estimate those probabilities which,
in order to account for their uncertainty,
are treated as random. We now describe
a general logistic regression approach for that purpose.

Assume that at time $t$, the sample of sites $S_t$ among the population of
$N$ sites is selected by a PPS (probability proportional
to size) sample survey design $u\in S_t$ being included
with probability $ \pi_{tu} $. That probability is assumed to
depend on all responses, both observed and
those unmeasured over the time
period $1\dvtx (t-1)$ (the latter being
treated as latent variables, analogous to the $S$'s in D10).
Thus, in terms of the measured and unmeasured
responses $Y$ and the vector of binary indicators of
selected/rejected sites $R$, the conditional distribution of the
probability of selection is
\begin{eqnarray}\label{eq:phi}
\operatorname{\mbox{logit}}[\pi_{tu}] &= & \operatorname{\mbox{logit}}\bigl[ P(R_{tu} = 1 \mid
\mathbf {y}_{1:(t-1)},\mathbf{r}_{1:(t-1)} ) \bigr]
\nonumber
\\[-8pt]
\\[-8pt]
\nonumber
& = & G(\mathbf{y}_{1:(t-1)},\mathbf{r}_{1:(t-1)} )
\end{eqnarray}
for some function $0\leq G \leq1$. That function is
our analogue of the preferential sampling intensity
in Assumption~2 in D10.

Under the assumption of the superpopulation model there will be
a predictive probability distribution for the
unmeasured responses. Values for these might be obtained using, for
example, geostatistical methods, which under repeated imputation will
allow $k=1,\ldots, K^*$ replicate data sets. Each replicate enables us
to fit $G$
through logistic regression to get
$\hat{G}^k, k=1,\ldots,K^*$ and in turn
$\hat{\pi}_{tu}^k, k=1,\ldots,K^*$.
This approach is analogous to equation (9)
in D10, with $R$ playing the role of $X$.

From these replicates, multiple values
of the HT estimator can be obtained, which allows an adjusted point
estimate of the population average to be found together with error
bands around the point estimate
to reflect the associated uncertainty.

The exact way in which these
are computed would depend on the scheme by
which the network was adapted as illustrated in
Section~\ref{sect:simulations}. We now discuss the case of a network
that expands monotonically over time, referred to as an ascending
staircase design. This forms the basis of one of the simulation studies
in Section~\ref{sect:simulations} together with one based on a
shrinking network.

%s4.1 #&#
\subsection{Expanding networks} Consider the case
of an ascending staircase design so that the network expands monotonically
over time [\citet{le2006statistical}], that is, $S_{(t-1)} \subseteq
S_t, t = 1,\ldots, T$, where
$S_0 $ is the null set and $S_T \subseteq{\cal D}$.
After a site, $u$, has entered the network it remains. Using an $^*$ to denote
unconditional probabilities, we assume
that initially sites in $S_1 \subset{\cal D} $ are selected
without replacement so that the selection probabilities are
$\pi_{1u}^* = \pi_{1u} = \mid S_1\mid/ N $, where in general
$\mid A \mid$ is the number of elements in a set $A$.
The HT estimator
for the population mean can now be calculated as
%
%e4.1 #&#
\begin{equation}
\label{eq:httime1} \hat{\mu}_1=\sum_{u\in S_1}
\frac{y_{1u} }{N\pi_{1u}^*}.
\end{equation}

At time 2, additional sites $ S_2\smallsetminus S_1$ must
be selected from ${\cal D} \smallsetminus S_1$ and
this is assumed to be done with probabilities
proportional to size at time 1, that is, based
on the $\{y_{1u}\}$ for these sites.
%%so that conditional probability of being selected at
%%$t=2$ given that $ u $ was not previously selected
%%is given by
%%\begin{equation}\label{eq:phiexample}
%%logit (\pi_{2u}) =\alpha_2 + \beta_2 y_{1u}.
%%\end{equation}
At time 2, the HT approach sees a single sample of sites
$S_2$. Inclusion in $S_2$ means a site was
either selected at time 1,
in which case it is certain to be in $S_2$,
or it was selected at time $t=2$ for the first time.
Hence,
overall for all sites in $u\in S_2$,
%
%e4.2 #&#
\begin{equation}
\label{eq:uncond2} \pi_{2u}^* = \pi_{1u}^* +\bigl(1-
\pi_{1u}^*\bigr)\pi_{2u}.
\end{equation}
However, at time 2, the $\pi$'s are unknown unlike those at time $1$
due to the unknown responses at time $1$
on which the preferential sampling was based and
the unknown monotone function of these responses
implicitly used by the Preferential Sampler.
The unknown responses
can be multiply imputed by standard
geostatistical (or other) methods to get
$\{\hat{y}_{1u}^k,  {\cal D} \smallsetminus S_1\}$
on replicate $k=1,\ldots,K$
%%%%%%%
%%%%%%
so that, in effect, we have a complete set of responses
over $\cald$.

Logistic regression can then be used to estimate the
probabilities of selection by fitting
%
%e4.3 #&#
\begin{equation}
\label{eq:phi2} \operatorname{\mbox{logit}} \bigl(\hat{\pi}_{2u}^k\bigr) =
\alpha_2 + \beta_2 \bigl[\hat{y}_{1u}^k
- \bar{\hat{y}}{}^k_{1\cdot}\bigr],
\end{equation}
%
%to the $N$ binary select-reject indicators for all sites
%$u\in\cald$, where $\hat{y}$ represents either observed
%or imputed values at time 1 as appropriate.
to the $N$ binary select-reject indicators for the set of
potentially new sites, \mbox{$u\in \cald\smallsetminus S_1$} where $\hat{y}$
represents either observed
or imputed values at time 1 as appropriate. In addition to sites in
$\cald\smallsetminus S_1$,
this model can be used to predict for sites in $ S_1$, allowing the
selection probabilities to be estimated for all $u\in \cald$ as required.
Replicate HT estimators at time $2$ are obtained by multiply
imputing the unknown responses and generating
multiple HT estimates:
%
%e4.4 #&#
\begin{equation}
\label{eq:httime2} \hat{\mu}_2^k=\sum
_{u\in S_2} \frac{y_{2u} }{N\hat{\pi}_{2u}^{*k}}.
\end{equation}
Their average yields the adjusted point estimator at time
$2$ and their empirical quantiles provide a means of estimating (95\%)
error bands
for the true population mean.

We can proceed in a similar fashion at time $3$. Here the required
imputation of the unobserved $y_{2u}$ is more complicated due to the
preferential sampling effect.
To do this, we adapt a key idea in \citet{pati2011bayesian}
and use logit transformations of the
estimated selection probabilities
as covariates that represent the preferential
selection effect. These can then be incorporated
in the spatial trend (mean field) model
for a Gaussian random
field superpopulation model.
Since at time $1$ we assume no preferential
selection, we let the required covariates be
$z_{1u} = \operatorname{\mbox{logit}} (\pi_{1u}), u \in{\cal D} \smallsetminus S_1$.
Then at time
$t=2$ we get
%
%e4.5 #&#
\begin{equation}
\label{eq:phiestimateexample} z_{2u} = \operatorname{\mbox{logit}}
 \bigl(\bar{\pi}_{2u}^{\cdot}
\bigr),\quad u \in{\cal D}
\end{equation}
by averaging the replicate values in
equation (\ref{eq:phi2}). %For completeness
%take $z_{2u} = z_{1u},  u \in S_1$.
Note that the $z$'s correspond
to what \citet{pati2011bayesian} call ``plug-in'' estimates.

%%In this way, all
%%$u \in S_T = \bigcup_{t=1}^T S_t \smallsetminus S_{t-1}$ have
%%an associated preferential sampling covariate, $S_0$ being the
%% null set. In other words a site $S_t \smallsetminus S_{t-1}$ has
%% for all times $ t^\prime\geq t$ the single covariate $Z_{t^\prime
%u}$
%% defined by $z_{t^\prime u} = z_{t u} $.
%%\subsubsection{The imputation}
%required in equation \ref{eq:phi} is done
%%by
We assume a Gaussian random field superpopulation model
with a Mat\'ern covariance matrix and spatial mean field
%
%e4.6 #&#
\begin{equation}
\label{eq:multnormal} E[Y_{2u} ] = \mu_{2u} + \vartheta
z_{2u},\qquad u \in{\cal D},
\end{equation}
where $\vartheta$ is an unknown regression
parameter.
This model is fit, including any unspecified parameters
in $\mu_{2u}$, using, for example, geostatistical methods
and by multiply imputing the unobserved values of the unobserved $\{
y_{2u}\}$.
Proceeding recursively in this way leads to a $K \times t $
matrix of replicates $\hat{\pi}_{tu}^{*k}$ for each $u$ at time $t$.
That in turn
yields replicates of the HT estimates for the
population mean at time~$t$.

Refinements of this approach would be possible,
including the addition of
a term in equation (\ref{eq:phiestimateexample}) to incorporate
spatial structure. This would be appropriate if it were known that
designers took spatial considerations into account, in coming
to their new-site admission decisions and in calculating their summary
statistics.
Model selection presents another challenge for equation
(\ref{eq:phi}). Selecting
appropriate predictors from the class of all possible
metrics that could be computed from previous exposure data will
be challenging.
Formal model selection approaches will generally
be impractical, necessitating reliance on some context specific
knowledge to help reduce the class of possibilities.

%s4.2 #&#
\subsection{Reducing networks}
The case in which a network is monotonically decreasing is somewhat
simpler, as the need for imputation of unmeasured responses is
eliminated. Starting from a set of $N$ sites, $S_1$ at time 1, then at
each time $t$, the set of sites that remain in the network will be $S_t
\subset S_{t-1}$. At each time point, the selection probabilities are
obtained from logistic regression and the HT estimates then constructed
based on equation (\ref{eq:estimator}).

The efficacy of the cases of decreasing and expanding networks are
explored using simulation in Section~\ref{sect:simulations}, with the
former also being the basis of the case study presented in Section~\ref
{sect:blacksmoke}.

%s5 #&#
\section{Simulation study}\label{sect:simulations}
This section describes simulation studies that explore
the approaches described in Section~\ref{sect:implementation}. Using
the terminology introduced in Section~\ref{sect:superpopmodel}, we
generate data for the underlying
environmental field (Nature) following the structure used in \citet
{gelfand2012effect}. Given this field, we simulate the role of the
Preferential Sampler who at each time $t$ selects sites for inclusion
at time $t+1$ with selection probabilities proportional to the
magnitude of the measurements (PPS). The
Statistician, having
only the measurements
at the selected
monitoring sites at time $t$, adopts
a superpopulation
model for all unmeasured responses and, knowing of the use of PPS, fits
a logistic regression model to the binary
site selection process for time $t+1$ to estimate
the site selection probabilities. The
HT estimator is then used to
adjust the annual average estimates for time $t+1$
for the effects of preferential sampling.

%%%%%%%%%%%%%%%%%%%%%%%%%%%%%%%%%%%
\subsection*{The underlying environmental field: Nature} To generate
emissions over space and time, we consider two cases from
\citet{gelfand2012effect}. The first of these considers emissions
arising from a point source (of pollution), while the second considers
pollution from three cities. Central to the approach advocated in
\citet
{gelfand2012effect} is that measured concentrations are based on
emissions, which in the case of the three cities are represented by
population densities. We generate data over 25 years for a finite
population of 1000 sites.

For the point source example, emissions are represented by a kernel
$x_u$ given by
%
%e5.1 #&#
\begin{equation}
x_u = \exp\bigl( -1.8 \Vert u - q \Vert\bigr),\qquad u \in{
\cal D}
\end{equation}
with $q = (0.25, 0.75)$.
The maximum and minimum values of $x$
are $1$ and $0.16$, respectively. The second scenario involves
three cities with centers located at $c_1 = (0.75, 0.75), c_2= (0.25,0.25),
c_3 = (0.75,0.25)$ and their
population densities given by
%
%e5.2 #&#
\begin{equation}
p_u = \exp\Bigl( -5 \min_i
\Vert u - c_i \Vert\Bigr), \qquad u \in{\cal D}.
\end{equation}
The maximum and minimum values of $p$ are $1$ and $0.019$, respectively.

The following adapts this approach to the cases considered here by
incorporating time. At each time $t$, we assume a linear relationship for
the mean concentrations and emissions.
For the two cases (point source, $\mathrm{M}_1$, and multi-city, $\mathrm{M}_2$),

\begin{longlist}[$\mathrm{M}_1$:]
\item[$\mathrm{M}_1$:] $\mu_{1tu} = \phi_1 x_{tu}$,
\item[$\mathrm{M}_2$:] $\mu_{2tu} = \phi_2 p_{tu}$.
\end{longlist}

The emission levels are assumed to decline
over time, $t = 1, \ldots, T, u \in{\cal D}$:
%
%e5.3 #&#
\begin{eqnarray}
\label{eq:x} x_{tu} &= & x_u - \gamma_{1u} (t
-1),
\\
\label{eq:p} p_{tu} & = & p_u - \gamma_{2u}
(t -1),
\end{eqnarray}
where decay parameters, $\gamma_{iu}$, are site specific; $\gamma_{iu}
= [a_i \mu_{iu} + b_i]\mu_{iu}$ for the two models $M_i$ [$a_1=
0.009391$, $b_1= 0.001216$, $a_2= 0.008156$, $b_2= 0.003686$---these values
were chosen to ensure that
after simulated Gaussian residuals are added to the spatial
mean, the simulated responses
will be nonnegative (with high probability) over all sites and
times]. Using this approach, the large mean values are greatly reduced
(50\% for the largest mean),
unlike the small ones (10\% for the smallest mean). In the following
simulations, the relationship between emissions and concentrations is
set to $\phi=2$. %%Table~\ref{table:simpars} gives the emission decay
%factors.

We select an irregular grid of 1000 points from a regular grid of
10,000 points to represent the population of possible sampling sites.
For each time $t$, the pollution field is simulated as a Gaussian field with
means $\mu_{jt}$ as described above and a fixed Mat\'ern spatial
correlation structure whose sill parameter differs for the two mean
models (again to ensure a high probability of nonnegative simulated values).

%
%
%t1 #&#
\begin{table}[b]
\caption{Spatial parameters used in the simulation studies}\label{table:simpars}
\begin{tabular*}{\textwidth}{@{\extracolsep{\fill}}lcccc@{}}
\hline
\textbf{Model} & \textbf{Nugget} & \textbf{Sill} & \textbf{Range} &
\multicolumn{1}{c@{}}{\textbf{Smoothness}} \\
\hline
$\mathrm{M}_1$ & 0 & 0.0079\phantom{0} & 0.5 & 0.5 \\
$\mathrm{M}_2$ & 0 & 0.00013 & 0.5 & 0.5 \\
\hline
\end{tabular*}
\end{table}

The correlation between the process responses
for sites separated
by a distance $d$ is given by
\[
\rho_{\theta}(d) = \frac{2}{2^{\kappa} \Gamma(\kappa) } \biggl( \frac{d\sqrt{2\kappa}}{\omega}
\biggr)^{\kappa} K_{\kappa} \biggl( \frac{d\sqrt{2\kappa}}{\omega} \biggr)
\]
for both models $M_i, i=1,2$. Here
$\kappa$ is the smoothness parameter, which in this case is $1/2$
to yield the exponential spatial correlation function. The $\omega$
determines the range of the model. The correlation would be
multiplied by the sill to get the spatial covariance function. The
values of these parameters are specified in
Table~\ref{table:simpars}. Thus, at each time point a random vector of
responses $Y_t^{1\times1000} \sim N_{1000}(\mu_t, \Sigma)$ is
generated, where $\Sigma$ is determined by the member of the
Mat\'ern family.
%%%%%%%%Oct2, 1013%%%%%%%%%%%%%%%%%%%%%%
The result: a matrix of simulated emission levels of dimension $1000
\times25$
which constitute the finite population to be studied in this section.
%%%%%%%%%%%%%%%%%%%%%%%%%%%%%%%%%%%

%%%%%%%%%%%%%%%%%%%%%%%%%%%%%%%%%%%%%%
\subsection*{Selection procedures: The preferential sampler}
We consider both of the cases described in the previous section: (i)
descending (network reducing in size) staircase and (ii) ascending
(network increasing) staircase of adaptive network design.

\begin{enumerate}[(ii)]
\item[(i)] Shrinking adaptive network design:

\begin{enumerate}[1.]
\i[1.] For time $t=1$, let $S_1 = {\cal D} $ be the entire population of
$N$ sites.
\i[2.] For each successive time $t=2,\ldots, T$, draw a sample $S_t \subset S_{t-1}$
of size $m_t = \mid S_t \mid= 25$ with sampling probabilities
proportional to size,
$\alpha+ \beta y_{tu}, u\in S_{t-1} $. For each of the two
emission scenarios, two cases are considered, the
first being that sampling is mildly preferential and the second
strongly preferential. The respective selection parameters are
given in Table~\ref{table:samplingparameters}.
\i[3.] Repeat step 2 1000 times to generate point estimates and 95\% error
bars for the annual estimates
the selections produce.
\end{enumerate}

%
%t2 #&#
\begin{table}
\tabcolsep=25pt
\caption{Parameters used to characterize
mild and strong preferential sampling for each of the two emission
scenarios considered in this simulation study}\label{table:samplingparameters}
\begin{tabular*}{\textwidth}{@{\extracolsep{\fill}}lcccc@{}}
\hline
&\multicolumn{2}{c}{\textbf{Mildly preferential}} &
\multicolumn{2}{c@{}}{\textbf{Strongly preferential}} \\[-6pt]
&\multicolumn{2}{c}{\hrulefill} &
\multicolumn{2}{c@{}}{\hrulefill} \\
\textbf{Scenario} & $\bolds{\alpha}$ & $\bolds{\beta}$ & $\bolds{\alpha}$ &
$ \bolds{\beta}$ \\
\hline
Point source &0.32\phantom{0} & 0.10\phantom{0}&0.32\phantom{0}&2\phantom{0.} \\
Multi-city &0.038 &0.010&0.038& 0.5\\
\hline
\end{tabular*}
\end{table}

\item[(ii)] Expanding adaptive network design:

\begin{enumerate}
\i[1.] Time $1$: Draw a simple random sample of $m_1 = 50$ sites $S_1$
without replacement from ${\cal D}$ of $N= 1000$ sites. Lack of
knowledge at that stage makes this selection model plausible.
\i[2.] Time $t = 2, \ldots, 25$: Draw a sample $S_t \supset S_{t-1}$
by adding an additional $m_t = 10$ sites selected (from the remaining
unselected sites from the original set of 1000 sites) with
probability proportional to size, again $\alpha+ \beta y_{tu}, u
\notin S_{t-1}$.
\i[3.] Repeat step 2 1000 times.
\end{enumerate}
\end{enumerate}

%%%%%%%%%%%%%%%%%%%%%%%%%%%%
\subsection*{Correcting for preferential sampling: The statistician}
This is done assuming a Gaussian random field (GRF) superpopulation
random response, that sampling probabilities are proportional to size
and the multivariate response vectors not autocorrelated. For clarity
of exposition, we drop the subscript $j$ denoting scenario model $M_j$.
The following are the steps needed for the two cases:

\begin{enumerate}[(i)]
\item[(i)] Shrinking adaptive network design:

\begin{enumerate}
\i[1.] Time $1$: $S_1 = \cald$.
\i[2.] Time $t=2,\ldots, T$: Use logistic regression as in
equation (\ref{eq:phi2}), using the sites in $S_1$
instead of $\cald \smallsetminus S_1$ from the expanding case,
%%model in Equation (\ref{eq:phiexample}) with $p=1$
to estimate
coefficients of the selection model ($\alpha$ and $\beta$) and hence
get an estimate of the conditional
selection probability $\hat{\pi}_{tu}$ for that time.
All relevant data are available so no prediction is needed.
\i[3.] Time $t=1,\ldots, T$: Compute the unconditional
site selection probabilities $\hat{\pi}_{tu}^*$ and the
HT estimate of the annual mean.
Note $ u \in S_t$ implies $ u \in S_{t^\prime},  t^\prime\leq t$ so
under the assumption of no autocorrelation
\[
\hat{\pi}_{tu}^*= \Pi_{t=1}^t \hat{
\pi}_{tu}.
\]
\i[4.] Repeat steps 2 and 3 to compute point estimates with 95\% error
bands for the
estimates.
\end{enumerate}
\end{enumerate}

\begin{figure}

\includegraphics{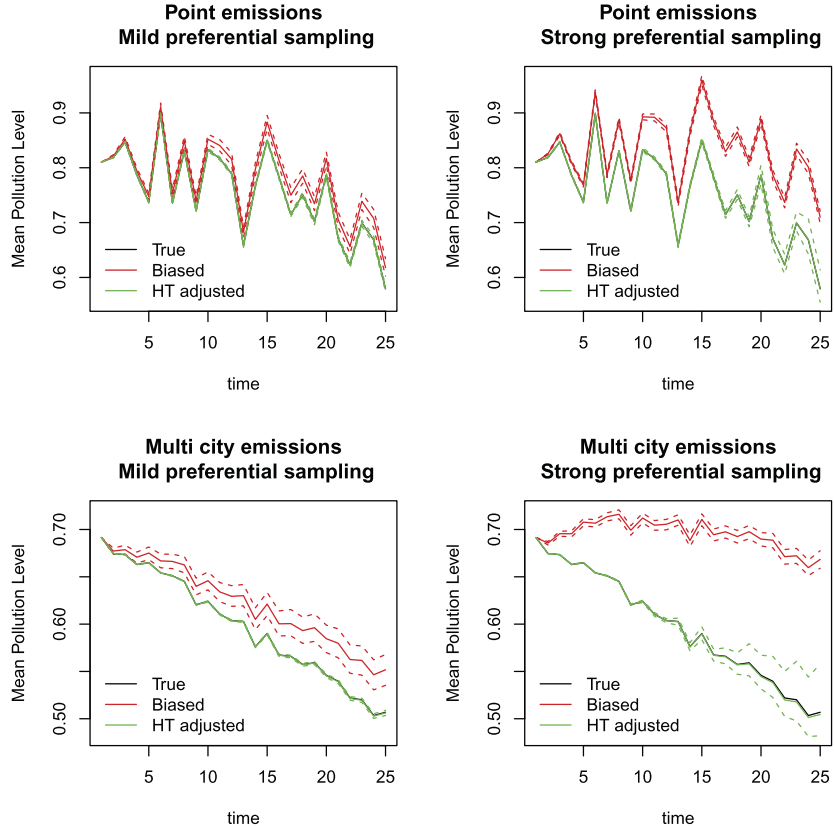}

\caption{Results from the simulation study of a network that is reduced
in size over time. Lines represent the population average over all
sites under the superpopulation models (black), the unadjusted
estimates (red) of that mean and the Horvitz--Thompson
adjusted estimates (green). Dotted lines show 95\% error bands based on
1000 simulated data sets. Upper and lower panels show results for
single and multiple point emission
scenarios, respectively. Note that for most of the times the black line
is not visible, as it is overlaid by the green line due to the closeness
of the adjusted estimates to the true values.}\vspace*{-5pt}
\label{fig:htadjustment}
\end{figure}

%%%Oct2_2013
Figure~\ref{fig:htadjustment} depicts the results for
four different cases. The top two panels are for
the case of a single emission source and the differing
levels of preferential sampling. The bottom
two show the corresponding results for the multi-city scenario.
In each case, the black
lines show the (true) average at each time $t$ over the
finite population of 1000 sites. The red lines give the (biased)
summaries for each time point together with an indication of the
variability over the multiple data set through the 95\% error bars (red
dotted lines). Green lines show the HT adjusted summaries at each time
point (with dotted lines signifying the associated variation). The
adjusted values are extremely close to the true values, a fact
reflected in the green lines overlaying the black (which are not
visible) in the upper panels.

%f1 #&#

%heere

The overall pattern in all cases is that levels decline over time,
reflecting a feature of
the superpopulation distribution, and error bands increase in width,
reflecting smaller samples.
When preferential sampling is weak (the two left-hand panels),
the red curve is relatively close to the black curve, although
significantly higher. The difference in the right-hand panels, when
preferential sampling is strong, is much more marked.
The HT adjustment improves the
estimators of the annual mean in all cases and that improvement is
dramatic with strong preferential sampling.

To provide a comparison with the estimates arising from applying the
method proposed here, we briefly consider the effects of two much
simpler approaches. Given the set of sites that were present at the
beginning of the period of interest, when a specific site ceased to be
part of the network, then the resulting missing data from that time
point might be imputed. A regression line might be fit to the available
measurements from that site and used to predict measurements from that
time. The average for each year would then be calculated using a
mixture of observed and predicted measurements for all sites at each
year. An even simpler approach might be to fill in missing data for
each site using the last recorded measurement, that is, filling in the
missing data with the most recent measurement available for that site.
The first approach would require a reasonable amount of data points for
each site (here chosen to be five) and would likely result in negative
predictions over the latter time years where the initial decline
observed at a site was strong. In such cases, predictions here are
truncated at zero. In both cases, applying these two simple approaches
resulted in the estimates of the yearly averages being overestimated in
all four of the cases presented in Figure~\ref{fig:htadjustment}. Filling in the missing
values using the last value proved to be consistently higher than the
true values, with the error increasing over time as might be expected.
In the examples shown in the panels of Figure~\ref{fig:htadjustment}, the overestimates in
the last (25th) year were 16\%, 14\%, 8\% and 11\% for the mild/strong point source and mild/strong multi-city cases, respectively.
The regression approach also consistently overestimated the yearly
averages, with the corresponding overestimates in the last year being
34\%, 18\%, 1\% and 7\%. In the third case (of mild preferential
sampling from multiple cities) the regression approach seems to do
well, as the downward slope of concentrations is close to linear,
however, this is not repeated when there is strong preferential
sampling and, as might be expected, applying these simple approaches
would result in summaries being overestimated, as would be seen when
just using the available data (black lines in Figure~\ref{fig:htadjustment}).

When the site selection probabilities are proportional to size, the HT
estimator has another desirable feature, that the $\{ y_{tu}/\pi_{tu}
\}
$ ratios will be nearer constant than if the $\{ \pi_{tu} \}$ are all
equal, as is the case with the unadjusted estimates [\citet
{stehman1994comparison}]. This will lead to a decrease in the
variability associated with the adjusted estimates compared to that for
the unadjusted ones, resulting in narrower 95\% error bars. In
practice, this will additionally be affected by factors such as the
strength of preferential sampling, it's cumulative effect over time and
the underlying variability of the population parameters. This can be
seen in Figure~\ref{fig:htadjustment} with the narrow error bars for
the adjusted estimates (compared to their unadjusted counterparts)
increasing in width over time and with the extent of the variability in
the underlying field.

For the second mode of sampling the overall setting is the same as the
first, but now the unobserved responses need to be imputed. A GRF
superpopulation model is assumed with Model $M_1$ being $Y_{jtu} = \nu
_{jtu} + \varepsilon_{jtu}$, where $\nu_{jtu} = \xi_1 x_{jtu}+ \xi_2
z_{tu}$. The structure of Model $M_2$ is the same with the $x$ replaced
by $p$ [as in equations~(\ref{eq:x}) and (\ref{eq:p})]. The
preferential sampling covariates $\{z_{tu}\}$ are specified below.
After fitting the $\{\xi_i, i=1,2\}$ and the parameters of the
covariance models, unobserved $y_{jtu}$ can be imputed
by standard geostatistical (or other) methods. Analysis proceeds as
follows for each of the two models:

\begin{enumerate}[(ii)]
\item[(ii)] Expanding adaptive network design:
\begin{enumerate}
\i[1.] Time $1$: Compute the HT estimate of the population mean $\hat{\mu
}_1$ using
equation (\ref{eq:httime1}). Impute the unobserved responses $\hat
{y}_{1u}, u \in\cald\smallsetminus S_1$.
\i[2.] Time $2$: Use the imputed responses at time 1
%$\hat{y}_{1u}$
and logistic regression as in equation (\ref{eq:phi2}) to estimate the
conditional selection probabilities $\hat{\pi}_{2u}, u \in S_2$. Then
estimate their unconditional probabilities of selection
$\hat{\pi}_{2u}^*$ using equation (\ref{eq:uncond2}).
%%%%%%%%%
%%heere
Thus, at time 2, for all $u \in S_2$, $\hat{\pi}_{2u}^* = \hat{\pi
}_{1u}^*+ (1 - \hat{\pi}_{1u}^*)
\hat{\pi}_{2u}$. Compute the HT estimate for each of the two scenario
models ($M_j, j=1,2$):
\[
\hat{\mu}_2 = \sum_{u \in S_2}
\frac{y_{2u}}{N\hat{\pi}_{2u}^*}.
\]
\i[3.] Time $3$: Compute $z_{2u}$ using equation (\ref
{eq:phiestimateexample}). Impute the unobserved $\hat{y}_{2u}, u \in
\cald\smallsetminus S_2$ as described above. Estimate the conditional
selection probabilities $\hat{\pi}_{3u}$ for sites in $S_3$ and then
their unconditional selection probabilities $\hat{\pi}_{3u}^*$. Compute
the HT estimate for each of the two scenario models ($M_j, j=1,2$):
\[
\hat{\mu}_2 = \sum_{u \in S_3}
\frac{y_{3u}}{N\hat{\pi}_{3u}^*}.
\]
\i[4.] Time $t = 4,\ldots,T$: Repeat step 3 after recursively updating it.
\i[5.] Repeat the previous steps for each of the 1000 replicate data sets,
at each time calculating the HT population average estimates along with
their 95\% error bands.
\end{enumerate}
\end{enumerate}

\begin{figure}

\includegraphics{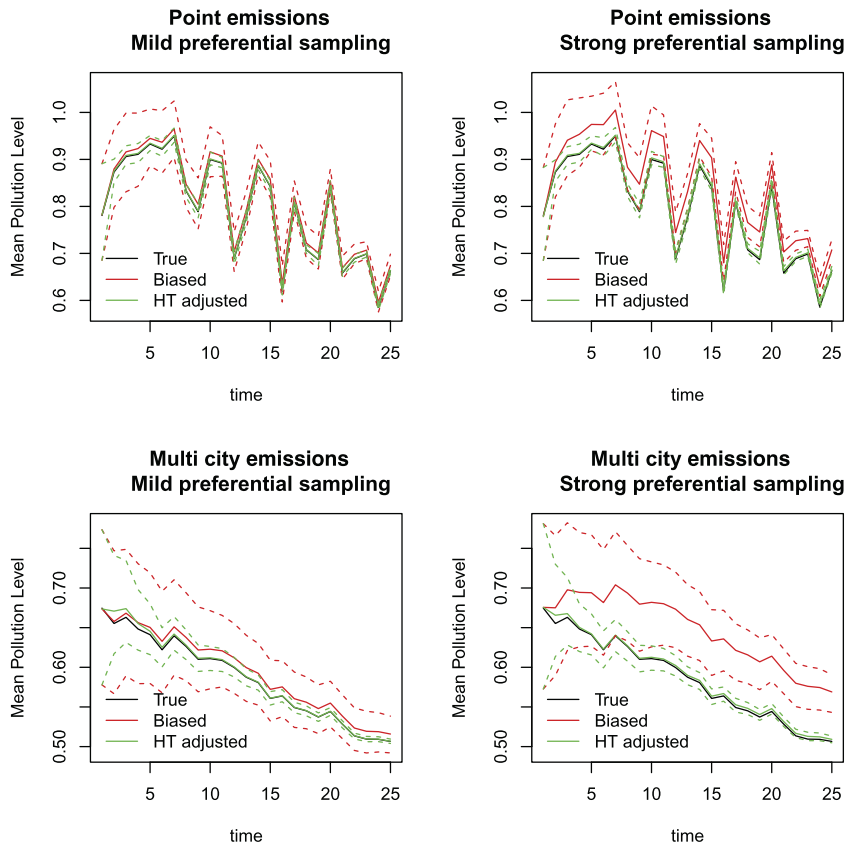}

\caption{Results from the simulation study of a network that is
increasing in size over time. Lines represent the population average
over all sites under the superpopulation models (black), the unadjusted
estimates (red) of that mean and the Horvitz--Thompson
adjusted estimates (green). Dotted lines show 95\% error bands based on
1000 simulated data sets. Upper and lower panels show results for
single and multiple point emission
scenarios, respectively. Note that the black line is often not visible,
as it is overlaid by the green line due to the closeness of the adjusted
estimates to the true values.}
\label{fig:htadjustment2}
\end{figure}

Figure~\ref{fig:htadjustment2} shows the corresponding results to those
seen in Figure~\ref{fig:htadjustment} for the expanding network case.
As with the deceasing network case, clear differences can be seen
between the true finite population means at each time point and those
estimated from the data arising from the preferential samples, with the
differences being greater in the case of strong preferential sampling.
This bias is again markedly reduced when using the HT adjusted
estimates. The width of the error bands for both
estimates are initially large due to the small sample sizes and they
decline in width as time goes on due to the increasing sample sizes. In
all cases
those for the HT estimator are the narrower of the two
for the reasons given above. Perhaps surprisingly, this reduction is observed
in all four cases over all
time, despite the need in this case to impute
unobserved responses in the case of the HT estimator.
Imputation does add uncertainty, however, as we see
in comparison to the widths seen in the
decreasing network example,
as would be expected given that these bands reflect
both variation due to the preferential resampling and
that associated with the imputation of unobserved responses.

%f2 #&#

%s6 #&#
\section{Case study: Black Smoke in the United Kingdom}\label{sect:blacksmoke}

In this case study we aim to address one of the paper's primary aims,
that of adjusting population level estimates of BS levels in the UK
over an extended period to make them unbiased. \citet{shaddick14}
provide evidence of preferential sampling in the reduction in the
network over time, and here we use the methods developed in
Sections \ref{sect:superpopmodel} and \ref{sect:implementation} and
demonstrated
in the simulation studies (Section~\ref{sect:simulations}) to adjust estimates of
overall average annual concentration levels as well as the number of
sites out of compliance.

We begin with a summary of the monitoring program for BS in Great
Britain. Although air pollution has been a concern for many centuries,
it became a global health issue in the early parts of the last century
after a number of high pollution episodes were linked to increased
health risks [\citet{firket:36}; \citet{ciocco61denora}; Ministry of
Public Health 1954]. As a result, attempts were made to measure air
pollution concentrations in a regular and systematic way.

Daily average BS has been shown to be a reasonable predictor of
PM$_{10}$. In general, PM$_{10}$ concentrations are usually higher than
BS except during high episodes and, hence, if BS exceeds the PM$_{10}$
limit, it is likely that PM$_{10}$ will also be out of compliance
[\citet
{muir1995black}].
Black smoke (BS) is one of a number of measures of particulate matter;
other examples include the coefficient of haze (CoH) and total
suspended particulates (TSP), as well as
direct measurements of PM$_{10}$ and PM$_{2.5}$. Each of these has been
associated with adverse health outcomes [for PM$_{10}$, \citet
{samet00particulate}; for PM$_{2.5}$, \citet{goldberg01pm2.5}; for TSP,
\citet{lee2010semi}; for BS, \citet{verhoeff96air}; for CoH, \citet
{gwynn00timeseries}].
Attempts have been made to standardize the
measures of pollution by converting the measurements into
``equivalent'' amounts of PM$_{10}$, for example, PM$_{10} \approx
0.55$ TSP, PM$_{10} \approx$ CoH/0.55, PM$_{10} \approx$ BS and
PM$_{10} \approx$ PM$_{2.5}$/0.6 [\citet{dockery:pope:94}].

In 1961 the world's first coordinated national air pollution monitoring
network was established in the UK using BS and sulphur dioxide
monitoring sites at around 1200 sites. As levels of BS
pollution have declined, the network has been progressively
rationalized, reduced, moved, replaced and by the mid-nineties it
comprised of ca. 200
sites.

The data on annual concentrations of BS ($\mu$gm$^{-3}$) used in this case study
were obtained from the UK National Air Quality Information
Archive. We use data from 1970--1996 and restrict to the case where
sites were withdrawn from the network over time. A small number of sites were
added during this period, but they are almost exclusively ones which
reported even higher concentrations, suggesting they were added
preferentially. For clarity, we consider only the reduction in the network.
We use the 624 sites that were operational in 1970 and which had at
least 5 measurements in the following 25 years, and these sites define
the finite population, that is, the concentrations measured at these
sites as characterizing the BS field over the UK.
For each year, $t$, data are available from $n_t$ sites, $t=1,\ldots,26$.
Measurements, $Z_{it}$, are the log of the annual means of the 24 hour
mean concentrations of BS divided by
a normalization constant to make them unitless (to be able to apply
logarithms). Over the study period, the number of sites was reduced
from $n_1=624$ to $n_{26}=193$ with the yearly means over all sites,
$\sum_{i=1}^{n_i} Z_{it}/n_i$, falling from 60.5 to 9.3 $\mu$mg$^{-3}$
over the same period.
%%%%%%%%%%%%%%%%%%%%%%%%
However, the preferential selection
used to reduce the network, and demonstrated in \citet{shaddick14},
suggests the latter number (the sample
average of the values of the surviving sites) is too high.
That calls for an adjustment of the form now described.
%%%%%%%%%%%%%%%%%%%%%%%%%

%f3 #&#
\begin{figure}%[b]

\includegraphics{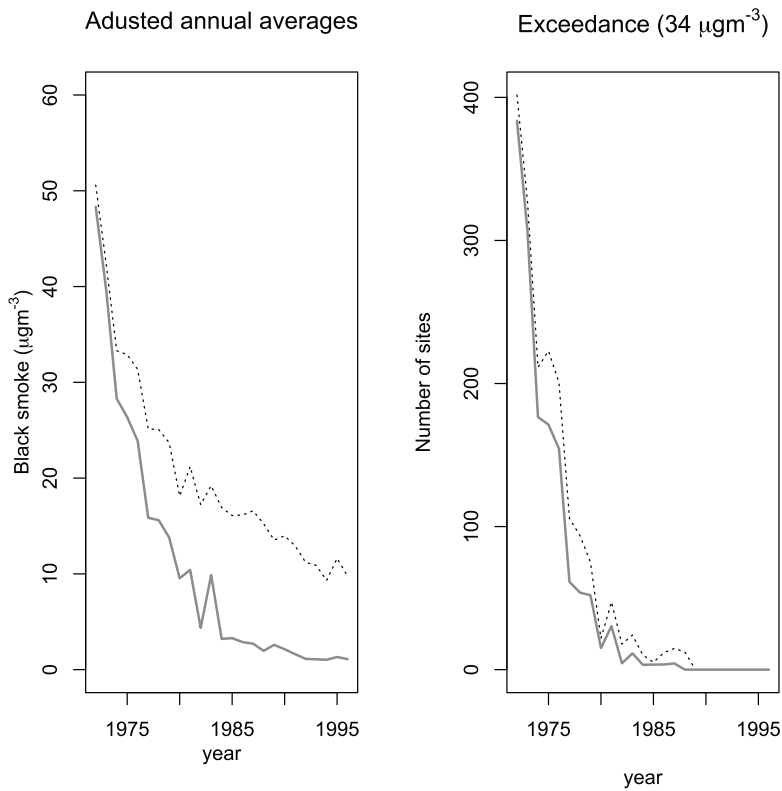}

\caption{Changes in the levels of black smoke within the UK from
1970--1996 and the effects of adjusting estimates of annual indices. The
left-hand panel shows the annual means over all sites (dotted black
line) together with adjusted values (solid grey line). The right-hand
panel shows the number of sites exceeding the EU guide value of 34 $\mu
$gm$^{-3}$ (dotted black line) together with the corrected values
(solid grey line).}
\label{log_res}
\end{figure}

Extensive analysis of these data suggests a log-Gaussian random field
superpopulation model [\citet{shaddick14}] for BS, because, in addition
to the desirable properties of
right-skew and nonnegativity, there is justification in terms of
the physical explanation of atmospheric chemistry [\citet{ott1990}]. We
therefore take the logarithms of BS concentrations after normalizing to
eliminate their units of measurement [\citet{monk11}], $Y_{it} = \log
(Z_{it}/78)$, where 78 $\mu$gm$^{-3}$ represents roughly the average
level of BS at the beginning of the time at which the network was operational.

Working on the log-scale described above, we applied
the methods described in Section~\ref{sect:simulations} for (i)
decreasing networks to adjust the annual arithmetic averages for the
effects of preferential sampling. Two characteristics associated with
the responses seem of natural interest. The first we consider is the
set of annual averages across these 624 sites, as these could be
published to show the effect of regulatory policy over time. The
left-hand panel of Figure~\ref{log_res} shows the estimated geometric
annual mean levels over time (dotted black line) together with the HT
adjusted ones (solid grey line). It clearly shows the adjustment
reduces the estimates of the average levels. Since the standard unit
for calculating relative risks of particulates in health effects
analysis is 10~$\mu$gm$^{-3}$, the difference seems important, being
more than one of these standard units over much of the period. This is
because, on the log-scale, responses with low values get high weights
in the HT adjusted arithmetic average, since their chances of making
the cut in every successive year, say, to 1985, for example, is very
small. As in Section~\ref{sect:simulations}, we also briefly consider the effects of two
very simple approaches to ``filling in'' the missing data after a site
has been excluded from the network: (i) using the last recorded value
throughout the following years and (ii) using prediction from linear
regression. Further details can be found in Section~\ref{sect:simulations}. As might be
expected, using the first approach here results in much higher
estimates of the annual averages, for example, in 1996 the estimate was
19.1 $\mu$gm$^{-3}$ compared to 9.8 $\mu$gm$^{-3}$ obtained from the
available data and 1.1 $\mu$gm$^{-3}$ using the HT approach. The
corresponding estimate using the linear regression approach was 2.4
$\mu
$gm$^{-3}$. While this appears a not unreasonable estimate, the
simplistic nature of the correction means that in the early years
of the analysis it produces estimates which are much higher than that
obtained from the available data and the HT estimates. In 1972, for
example the estimated annual average would be 60.5 $\mu$gm$^{-3}$ using
linear regression compared to 50.6 $\mu$gm$^{-3}$ using the available
data and 48 $\mu$gm$^{-3}$ using the HT approach, respectively.

%heere

The second characteristic we consider is potentially of even greater
operational importance, the number of sites in nonattainment, that is,
those which do not comply with the air quality standards in a given
year. This number is a surrogate for the cost of mitigation for putting
the BS concentrations into compliance. For example, as part of the
analysis of the impact of the various ozone standards considered by the
EPA's CASAC Ozone Committee in 2008, the EPA Staff predicted the
fraction of monitored counties in the United States that would be out
of compliance. For the standards that were finally proposed by the
Committee, that percent was found to be 86\%. Although the US Clean Air
Act (\surl{epa.gov/oar/caa/title1.html}) of 1970, under whose mandate the
CASAC was created, rules out economic impact in consideration of
standards designed to protect public health, nevertheless, policy
making cannot ignore the cost of attainment which can be substantial.
The right-hand panel of Figure~\ref{log_res} shows the number
of sites each year that exceed the 1980 EU guide value of 34 $\mu$gm$^{-3}$
[\citet{EC1980}]. The dotted black line is the number of exceeding sites
based on the recorded data, with the solid grey line the numbers after
adjustment for the preferential sampling. The unadjusted numbers are
the fraction in the monitoring network out of compliance multiplied by
the finite population total of $N=624$. Their adjusted counterparts are
found by applying the HT weights to the 1s present in the summation
used to calculate that fraction. Table~\ref{tab_leg} shows the number
of sites exceeding the EU limit of 68~$\mu$gm$^{-3}$ and the guide
values of 51 and 34 $\mu$gm$^{-3}$, where a reduction of the number of
exceedances can be seen when adjusting for the preferential sampling.
For example, in 1974, the crude estimate gives 211 of the 624 sites out
of compliance with the 34~$\mu$gm$^{-3}$ criterion, while its adjusted
counterpart is just 189. This is a substantial difference,
given the economic cost of mitigation. Note the large number of zeros
reflect the decline over time in the levels of BS.

Overall we see quite a substantial overestimation of important finite
population parameters due to the preferential sampling in published estimates.

%t3 #&#
\begin{table}
\caption{Estimated number of sites exceeding regulatory guide values
for black smoke, with and without adjustment for preferential sampling
using Horivitz--Thompson estimators}
\label{tab_leg}
\begin{tabular*}{\textwidth}{@{\extracolsep{\fill}}lcccccc@{}}
\hline
&\multicolumn{2}{c}{\textbf{Limit 68} $\bolds{\mu}$\textbf{gm}$\bolds{^{-3}}$} &
\multicolumn{2}{c}{\textbf{Guide 51} $\bolds{\mu}$\textbf{gm}$\bolds{^{-3}}$} &
\multicolumn{2}{c@{}}{\textbf{Guide 34} $\bolds{\mu}$\textbf{gm}$\bolds{^{-3}}$} \\[-6pt]
&\multicolumn{2}{c}{\hrulefill} &
\multicolumn{2}{c}{\hrulefill} &
\multicolumn{2}{c@{}}{\hrulefill} \\
& \textbf{Unadjusted} & \textbf{Adjusted} & \textbf{Unadjusted} & \textbf{Adjusted} &
\textbf{Unadjusted} &
\multicolumn{1}{c@{}}{\textbf{Adjusted}} \\
\hline
1972 & 129 & 123 & 236 & 225 & 402 & 384 \\
1973 & \phantom{0}73 & \phantom{0}68 & 153 & 143 & 327 & 306 \\
1974 & \phantom{0}31 & \phantom{0}28 & \phantom{0}94 & \phantom{0}84 & 211 & 189 \\
1975 & \phantom{0}21 & \phantom{0}19 & \phantom{0}58 & \phantom{0}52 & 223 & 201 \\
1976 & \phantom{0}19 & \phantom{0}18 & \phantom{0}50 & \phantom{0}47 & 201 & 189 \\
1977 & \phantom{00}7 & \phantom{00}6 & \phantom{0}23 & \phantom{0}20 & 106 & \phantom{0}94 \\
1978 & \phantom{00}7 & \phantom{00}7 & \phantom{0}18 & \phantom{0}17 & \phantom{0}94 & \phantom{0}87 \\
1979 & \phantom{00}8 & \phantom{00}7 & \phantom{0}21 & \phantom{0}19 & \phantom{0}75 & \phantom{0}71 \\
1980 & \phantom{00}0 & \phantom{00}0 & \phantom{00}6 & \phantom{00}6 & \phantom{0}22 & \phantom{0}21 \\
1981 & \phantom{00}2 & \phantom{00}2 & \phantom{0}11 & \phantom{0}10 & \phantom{0}47 & \phantom{0}42 \\
1982 & \phantom{00}0 & \phantom{00}0 & \phantom{00}0 & \phantom{00}0 & \phantom{0}18 & \phantom{00}9 \\
1983 & \phantom{00}0 & \phantom{00}0 & \phantom{0}10 & \phantom{00}5 & \phantom{0}24 & \phantom{0}16 \\
1984 & \phantom{00}0 & \phantom{00}0 & \phantom{00}0 & \phantom{00}0 & \phantom{0}10 & \phantom{00}8 \\
1985 & \phantom{00}0 & \phantom{00}0 & \phantom{00}0 & \phantom{00}0 & \phantom{00}5 & \phantom{00}4 \\
1986 & \phantom{00}0 & \phantom{00}0 & \phantom{00}0 & \phantom{00}0 & \phantom{0}12 & \phantom{00}6 \\
1987 & \phantom{00}0 & \phantom{00}0 & \phantom{00}0 & \phantom{00}0 & \phantom{0}15 & \phantom{00}7 \\
1988 & \phantom{00}0 & \phantom{00}0 & \phantom{00}0 & \phantom{00}0 & \phantom{0}12 & \phantom{00}3 \\
1989 & \phantom{00}0 & \phantom{00}0 & \phantom{00}0 & \phantom{00}0 & \phantom{00}0 & \phantom{00}0 \\
$\vdots$ & \phantom{00}$\vdots$ & \phantom{00}$\vdots$ & \phantom{00}$\vdots$ & \phantom{00}$\vdots$ & \phantom{00}$\vdots$ &
\phantom{00}$\vdots$ \\
1996 & \phantom{00}0 & \phantom{00}0 & \phantom{00}0 & \phantom{00}0 & \phantom{00}0 & \phantom{00}0 \\
\hline
\end{tabular*}
\end{table}

%s7 #&#
\section{Discussion}\label{sect:conclusions}

A number of methods have been discussed in this paper for modeling the
probability of selection in preferential sampling and we have developed
a general framework using a superpopulation modeling approach. Taking a
public policy perspective,
we have emphasized the HT approach to mitigating the effects of
preferential sampling in order to get unbiased estimators.

Having the space--time series of sites in Section~\ref{sect:blacksmoke}
enables us to do something that is not possible with only spatial data,
namely, to study the preferential selection process itself. This is
performed using logistic regression to estimate the selection
probabilities (Sections \ref{sect:simulations} and \ref
{sect:blacksmoke}). The results from the simulation studies in
Section~\ref{sect:simulations} suggest
the method proposed in this paper compensates for the effect of
preferentially sampling
and reduces bias. Section~\ref{sect:blacksmoke} shows that this
adjustment can be substantial in practice.

The case study in Section~\ref{sect:blacksmoke} demonstrated the use of
the method where the parameters being estimated are numerical features
of the finite population of exposures of air pollution. The finite
population in this case were monitoring sites in the UK that were
measuring BS in 1970. From that time there was a dramatic reduction in
the size of the monitoring network and subsequently only subsets of
those exposures were measured. In the case study we apply the methods
we have developed to adjust these estimates for the effects of
preferential sampling. The results show reductions in the estimates,
illustrating how the preferential siting of monitors where exposures
are high gives an exaggerated impression of the level of BS and the
number of sites that are in noncompliance. Note that selection bias can
accumulate over time and can require increasingly greater adjustments.
Although the effect on estimates of the number of sites out of
compliance appears not to be as dramatic, it is substantial, especially
considering that forcing attainment of standards can entail large costs.

In the case study, we assumed a log-Gaussian superpopulation model for
BS and so the geometric, rather than arithmetic annual average, must be
used to characterize the finite population's annual mean level. This
metric is often used for particulate air pollution; see, for example,
\citet{mueller1994characterization}. In Section~\ref
{sect:superpopmodel} (case 2) we show that in such cases
spatial-correlation-adjusted (unequal) weights ought to be used to
characterize the finite population mean levels. However, this is
extremely unlikely to be the case in reality and, as a concession to
standard practice, we use equal weights, leaving this issue for
exploration in future work. Asymptotic theory and variance
approximations are available for the Horvitz--Thompson estimator, which
would enable approximate error bands to be computed for the adjusted
estimates. However, these would not include the additional uncertainty
associated with the estimation of the selection probabilities
themselves. So we plan in future work to develop an approach that would
also involve the construction of measures of uncertainty associated
with the sampling weights and, consequently, the adjusted annual
averages and exceedances for a single data analysis (in the simulation
study 95\% error bars are obtained by repeated simulation). If both the
modeling of the weights and the adjustment process were combined within
a Bayesian framework, it may be possible to propagate the uncertainty
in the estimation of the weights (including that which may arise due to
spatial prediction in the case of expanding networks) through to the
adjusted values.

The HT approach will not always be the most appropriate approach. In
some cases a likelihood-based approach may be feasible provided that
the preferential sampling can be modeled. That could be the case in the
context of air pollution and health in epidemiological analyses, for,
as \citet{guttorp:2010} point out, the air pollution monitoring sites
may be intentionally located for reasons such as the need to measure
the following: (i) background levels outside of urban areas; (ii)
levels in residential areas; and (iii) levels near pollutant sources.
Then the method in D10 may be useful.
A possible alternative approach to the method in D10 is described in
\citet{shaddick2012unbiasing}. While it resembles the point process
model, it is designed for discrete site domains as seen in this paper.
Moreover, being based on intensities for paralyzable particle counters,
it would allow for preferential sampling designs with varying
intensities over time.

The approach taken in the paper will work best when at least some
representatives of the general population of sites continue to be
sampled since the selection weights would then be able to compensate
for their underrepresentation in computing population statistics. In
the case study, the BS network was originally set up to try and provide
monitoring for a cross-section of expected cities and pollution
levels, although in reality the original set of locations would have
included some form of preferential sampling. In the absence of such
representation or good background knowledge of how the biased selection
was made, there would seem to be no alternative but to augment the
network with some possibly temporary monitors. That leads to a design
problem about the optimal selection of those sites. In this case, then
we would be unbiasing the design rather than unbiasing the estimates,
and that would need a different approach than that described in this paper.

The analyses reported in this paper and its predecessors [\citet
{diggle2010geostatistical},
\citet{gelfand2012effect}, \citet{pati2011bayesian}], coupled with the
importance and widespread use of environmental monitoring networks,
points to the need for further exploration and confirmation of the
results of these analyses. We recognize there are limitations in the
preferential sampling models used in the simulation studies in these
papers. In practice, site selection is complex, involving committees,
guidelines and negotiations, and local administrators in affluent areas
demanding a monitor in their municipality. However, they do
convincingly demonstrate that preferential site selection does have
serious adverse consequences, something that does not appear to have
been recognized by agencies charged with formulating regulatory
guidelines. Network data are commonly used as if they represent a true
reflection of underlying environmental fields. Of course, there are
occasions when sites must be preferentially located, for example, to
check adherence to standards around industrial sources. In such cases
it may be undesirable for the data to be used for other, possibly
unintended, purposes, but if it is so used,
it should be adjusted for the possible effects of preferential sampling
using methods such as that presented in this paper.

\section*{Acknowledgments} The genesis of the work described in this
paper was
the 2009--2010 thematic program on spatial analysis for environmental
mapping, epidemiology and climate change developed by the Statistical
and Applied Mathematics Institute on spatial statistics. We are
grateful to members of its Working Group on preferential sampling, of
which the second author was a member, for discussions that stimulated
our interest in that topic. We would like to thank Yi Liu for his
assistance in the comparison of methods. We would also like to offer
great thanks to the Editor, Associate Editor and anonymous referees who
offered valuable suggestions which have greatly improved the paper, one
of which offered an extraordinarily detailed list of comments and
corrections without which the revision of the manuscript would have
been a much greater task.

%%\appendix
%%\section{Simulation code}

% imsref loaded by akundreckaite, 2014-06-23 13:28:35
% imsref loaded by akundreckaite, 2014-06-23 13:53:20
%

% zodis "Acknowledgments" paliekamas pagal autoriu

%suskaldyti doi

\printaddresses

\begin{thebibliography}{40}

%b1 #&#
\bibitem[\protect\citeauthoryear{Ainslie, Reuten and
Steyn}{2009}]{ainslieapplication2009}
%
\begin{barticle}[auto:STB|2014/06/18|12:29:53]
\bauthor{\bsnm{Ainslie},~\bfnm{B.}\binits{B.}},
\bauthor{\bsnm{Reuten},~\bfnm{C.}\binits{C.}},
\bauthor{\bsnm{Steyn},~\bfnm{D.~G.}\binits{D.~G.}},
\bauthor{\bsnm{Le},~\bfnm{N.~D.}\binits{N.~D.}}
\AND
\bauthor{\bsnm{Zidek},~\bfnm{J.~V.}\binits{J.~V.}}
(\byear{2009}).
\btitle{Application of an entropy-based Bayesian optimization
technique to the redesign of an existing monitoring network for single
air pollutants}.
\bjournal{Journal of Environmental Management}
\bvolume{90}
\bpages{2715--2729}.
\end{barticle}
%
\bptok{imsref}%
% NOT OUTPUTED:
% number = 8
\endbibitem

%b2 #&#
\bibitem[\protect\citeauthoryear{Binder}{1983}]{binder1983variances}
%
\begin{barticle}[mr]
\bauthor{\bsnm{Binder},~\bfnm{David~A.}\binits{D.~A.}}
(\byear{1983}).
\btitle{On the variances of asymptotically normal estimators from
complex surveys}.
\bjournal{Internat. Statist. Rev.}
\bvolume{51}
\bpages{279--292}.
\bid{doi={10.2307/1402588}, issn={0306-7734}, mr={0731144}}
\end{barticle}
%
\bptok{imsref}%
% NOT OUTPUTED:
% issn = 0306-7734
% url = http://dx.doi.org/10.2307/1402588
% number = 3
% coden = ISTRDP
% fjournal = International Statistical Review. Revue International de
%Statistique
\endbibitem

%b3 #&#
\bibitem[\protect\citeauthoryear{Binder and Patak}{1994}]{binder94}
%
\begin{barticle}[mr]
\bauthor{\bsnm{Binder},~\bfnm{David~A.}\binits{D.~A.}} \AND
\bauthor{\bsnm{Patak},~\bfnm{Zdenek}\binits{Z.}}
(\byear{1994}).
\btitle{Use of estimating functions for estimation from complex surveys}.
\bjournal{J. Amer. Statist. Assoc.}
\bvolume{89}
\bpages{1035--1043}.
\bid{issn={0162-1459}, mr={1294748}}
\end{barticle}
%
\bptok{imsref}%
% NOT OUTPUTED:
% issn = 0162-1459
% url =
%%http://links.jstor.org/sici?sici=0162-1459(199409)89:427<1035:UOEFFE>2.0.CO;2-B&origin=MSN
% number = 427
% coden = JSTNAL
% fjournal = Journal of the American Statistical Association
\endbibitem

%b4 #&#
\bibitem[\protect\citeauthoryear{Chang et~al.}{2007}]{chang2007designing}
%
\begin{barticle}[mr]
\bauthor{\bsnm{Chang},~\bfnm{Howard}\binits{H.}},
\bauthor{\bsnm{Fu},~\bfnm{Audrey~Qiuyan}\binits{A.~Q.}},
\bauthor{\bsnm{Le},~\bfnm{Nhu~D.}\binits{N.~D.}} \AND
\bauthor{\bsnm{Zidek},~\bfnm{James~V.}\binits{J.~V.}}
(\byear{2007}).
\btitle{Designing environmental monitoring networks to measure extremes}.
\bjournal{Environ. Ecol. Stat.}
\bvolume{14}
\bpages{301--321}.
\bid{doi={10.1007/s10651-007-0020-5}, issn={1352-8505}, mr={2405332}}
\end{barticle}
%
\bptok{imsref}%
% NOT OUTPUTED:
% issn = 1352-8505
% url = http://dx.doi.org/10.1007/s10651-007-0020-5
% number = 3
% coden = EESTFM
% fjournal = Environmental and Ecological Statistics
\endbibitem

%b5 #&#
\bibitem[\protect\citeauthoryear{Cicchitelli and
Montanari}{2012}]{cicchitelli12}
%
\begin{barticle}[mr]
\bauthor{\bsnm{Cicchitelli},~\bfnm{Giuseppe}\binits{G.}} \AND
\bauthor{\bsnm{Montanari},~\bfnm{Giorgio~E.}\binits{G.~E.}}
(\byear{2012}).
\btitle{Model-assisted estimation of a spatial population mean}.
\bjournal{Internat. Statist. Rev.}
\bvolume{80}
\bpages{111--126}.
\bid{doi={10.1111/j.1751-5823.2011.00164.x}, issn={0306-7734}, mr={2990348}}
\end{barticle}
%
\bptok{imsref}%
% NOT OUTPUTED:
% issn = 0306-7734
% url = http://dx.doi.org/10.1111/j.1751-5823.2011.00164.x
% number = 1
% fjournal = International Statistical Review. Revue Internationale de
%Statistique
\endbibitem

%b6 #&#
\bibitem[\protect\citeauthoryear{Ciocco and Thompson}{1961}]{ciocco61denora}
%
\begin{barticle}[auto:STB|2014/06/18|12:29:53]
\bauthor{\bsnm{Ciocco},~\bfnm{A.}\binits{A.}} \AND
\bauthor{\bsnm{Thompson},~\bfnm{D.~J.}\binits{D.~J.}}
(\byear{1961}).
\btitle{A follow-up of donora ten years after: Methodology and findings}.
\bjournal{Am. J. Public Health Nations Health}
\bvolume{51}
\bpages{155--164}.
\end{barticle}
%
\bptok{imsref}%
\endbibitem

%b7 #&#
\bibitem[\protect\citeauthoryear{Dawid}{2010}]{dawid:2010}
%
\begin{barticle}[auto]
\bauthor{\bsnm{Dawid},~\bfnm{P.}\binits{P.}}
(\byear{2010}).
\btitle{Discussion of ``Geostatistical inference under preferential sampling''
by Diggle, P.~J., Menezes, R. and Su, T}.
\bjournal{J. R. Stat. Soc. Ser. C. Appl. Stat.}
\bvolume{59}
\bpages{191--232}.
%mr={2744471}}
\bptnote{check related}%
\end{barticle}
%
\bptok{imsref}%
% NOT OUTPUTED:
% issn = 0035-9254
% url = http://dx.doi.org/10.1111/j.1467-9876.2009.00701.x
% number = 2
% fjournal = Journal of the Royal Statistical Society. Series C.
%Applied Statistics
\endbibitem

%b8 #&#
\bibitem[\protect\citeauthoryear{Diggle, Menezes and
Su}{2010b}]{diggle2010geostatistical}
%
\begin{barticle}[mr]
\bauthor{\bsnm{Diggle},~\bfnm{Peter~J.}\binits{P.~J.}},
\bauthor{\bsnm{Menezes},~\bfnm{Raquel}\binits{R.}} \AND
\bauthor{\bsnm{Su},~\bfnm{Ting-li}\binits{T.-l.}}
(\byear{2010}b).
\btitle{Geostatistical inference under preferential sampling}.
\bjournal{J. R. Stat. Soc. Ser. C. Appl. Stat.}
\bvolume{59}
\bpages{191--232}.
\bid{doi={10.1111/j.1467-9876.2009.00701.x}, issn={0035-9254}, mr={2744471}}
\end{barticle}
%
\bptok{imsref}%
% NOT OUTPUTED:
% issn = 0035-9254
% url = http://dx.doi.org/10.1111/j.1467-9876.2009.00701.x
% number = 2
% fjournal = Journal of the Royal Statistical Society. Series C.
%Applied Statistics
\endbibitem

%b9 #&#
\bibitem[\protect\citeauthoryear{Dockery and Pope CA III}{1994}]{dockery:pope:94}
%
\begin{barticle}[auto:STB|2014/06/18|12:29:53]
\bauthor{\bsnm{Dockery},~\bfnm{D.}\binits{D.}}
\and
\bauthor{\bsnm{Pope CA III}}
(\byear{1994}).
\btitle{Acute respiratory effects of particulate air pollution}.
\bjournal{Annu. Rev. Public Health}
\bvolume{15}
\bpages{107--132}.
\end{barticle}
%
\bptok{imsref}%
\endbibitem

%b10 #&#
\bibitem[\protect\citeauthoryear{EPA}{2006}]{ozone05}
%
\begin{bmisc}[auto:STB|2014/06/18|12:29:53]
\borganization{EPA}
(\byear{2006}).
\bhowpublished{Air quality criteria for ozone and related
photochemical oxidants. EPA/600/R-05/004aF-cF.}
\end{bmisc}
%
\bptok{imsref}%
% NOT OUTPUTED:
% sortkey = EPA(2006
% howpublished =.
\endbibitem

%b11 #&#
\bibitem[\protect\citeauthoryear{EPA}{2009}]{usepa2009national}
%
\begin{bmisc}[auto:STB|2014/06/18|12:29:53]
\borganization{EPA}
(\byear{2009}).
\bhowpublished{National Lakes Assessment: A collaborative survey of
the nation's lakes. EPA 841-R-09-001.}
\end{bmisc}
%
\bptok{imsref}%
% NOT OUTPUTED:
% sortkey = EPA(2009
% howpublished = EPA.
\endbibitem

%b12 #&#
\bibitem[\protect\citeauthoryear{European Commision}{1980}]{EC1980}
%
\begin{bmisc}[auto:STB|2014/06/18|12:29:53]
\borganization{European Commision}
(\byear{1980}).
\bhowpublished{Council directive 80/779/{EEC} of 15 July 1980 on air
quality limit values and guide values for sulphur dioxide and suspended
particulates.}
\end{bmisc}
%
\bptok{imsref}%
% NOT OUTPUTED:
% sortkey = European(1980
% howpublished =.
\endbibitem

%b13 #&#
\bibitem[\protect\citeauthoryear{Firket}{1936}]{firket:36}
%
\begin{barticle}[auto:STB|2014/06/18|12:29:53]
\bauthor{\bsnm{Firket},~\bfnm{J.}\binits{J.}}
(\byear{1936}).
\btitle{Fog along the Meuse valley}.
\bjournal{Trans. Faraday Soc.}
\bvolume{32}
\bpages{1191--1194}.
\end{barticle}
%
\bptok{imsref}%
\endbibitem

%b14 #&#
\bibitem[\protect\citeauthoryear{Gelfand, Sahu and
Holland}{2012}]{gelfand2012effect}
%
\begin{barticle}[mr]
\bauthor{\bsnm{Gelfand},~\bfnm{Alan~E.}\binits{A.~E.}},
\bauthor{\bsnm{Sahu},~\bfnm{Sujit~K.}\binits{S.~K.}} \AND
\bauthor{\bsnm{Holland},~\bfnm{David~M.}\binits{D.~M.}}
(\byear{2012}).
\btitle{On the effect of preferential sampling in spatial prediction}.
\bjournal{Environmetrics}
\bvolume{23}
\bpages{565--578}.
\bid{doi={10.1002/env.2169}, issn={1180-4009}, mr={3020075}}
\end{barticle}
%
\bptok{imsref}%
% NOT OUTPUTED:
% issn = 1180-4009
% url = http://dx.doi.org/10.1002/env.2169
% number = 7
% fjournal = Environmetrics
\endbibitem

%b15 #&#
\bibitem[\protect\citeauthoryear{Godambe and
Thompson}{1986}]{godambe1986parameters}
%
\begin{barticle}[mr]
\bauthor{\bsnm{Godambe},~\bfnm{V.~P.}\binits{V.~P.}} \AND
\bauthor{\bsnm{Thompson},~\bfnm{M.~E.}\binits{M.~E.}}
(\byear{1986}).
\btitle{Parameters of superpopulation and survey population: Their
relationships and estimation}.
\bjournal{Internat. Statist. Rev.}
\bvolume{54}
\bpages{127--138}.
\bid{doi={10.2307/1403139}, issn={0306-7734}, mr={0962931}}
\end{barticle}
%
\bptok{imsref}%
% NOT OUTPUTED:
% issn = 0306-7734
% url = http://dx.doi.org/10.2307/1403139
% number = 2
% coden = ISTRDP
% fjournal = International Statistical Review. Revue International de
%Statistique
\endbibitem

%b16 #&#
\bibitem[\protect\citeauthoryear{Goldberg et~al.}{2001}]{goldberg01pm2.5}
%
\begin{barticle}[auto:STB|2014/06/18|12:29:53]
\bauthor{\bsnm{Goldberg},~\bfnm{M.~S.}\binits{M.~S.}},
\bauthor{\bsnm{Burnett},~\bfnm{R.~T.}\binits{R.~T.}},
\bauthor{\bsnm{Bailar 3rd},~\bfnm{J.~C.}\binits{J.~C.}},
\bauthor{\bsnm{Tamblyn},~\bfnm{R.}\binits{R.}},
\bauthor{\bsnm{Ernst},~\bfnm{P.}\binits{P.}},
\bauthor{\bsnm{Flegel},~\bfnm{J.}\binits{J.}},
\bauthor{\bsnm{Brook},~\bfnm{K.}\binits{K.}},
\bauthor{\bsnm{Bonvalot},~\bfnm{Y.}\binits{Y.}},
\bauthor{\bsnm{Singh},~\bfnm{R.}\binits{R.}},
\bauthor{\bsnm{Valois},~\bfnm{M.~F.}\binits{M.~F.}} \AND
\bauthor{\bsnm{Vincent},~\bfnm{R.}\binits{R.}}
(\byear{2001}).
\btitle{Identification of persons with cardiorespiratory conditions
who are at risk of dying from the acute effects of ambient air particles}.
\bjournal{Environ. Health Perspect}
\bvolume{109}
\bpages{487--494}.
\end{barticle}
%
\bptok{imsref}%
% NOT OUTPUTED:
% sortkey = Goldberg(2001
% howpublished =.
\endbibitem

%b17 #&#
\bibitem[\protect\citeauthoryear{Guttorp and Sampson}{2010}]{guttorp:2010}
%
\begin{barticle}[auto]
\bauthor{\bsnm{Guttorp},~\bfnm{P.}\binits{P.}}
\AND
\bauthor{\bsnm{Sampson},~\bfnm{P.}\binits{P.}}
(\byear{2010}).
\btitle{Discussion of Geostatistical inference under preferential sampling
by Diggle, P.~J., Menezes, R. and Su, T}.
\bjournal{J. R. Stat. Soc. Ser. C. Appl. Stat.}
\bvolume{59}
\bpages{191--232}.
%mr={2744471}}
\end{barticle}
%
\bptok{imsref}%
% NOT OUTPUTED:
% issn = 0035-9254
% url = http://dx.doi.org/10.1111/j.1467-9876.2009.00701.x
% number = 2
% fjournal = Journal of the Royal Statistical Society. Series C.
%Applied Statistics
\endbibitem

%b18 #&#
\bibitem[\protect\citeauthoryear{Gwynn, Burnett and
Thurston}{2000}]{gwynn00timeseries}
%
\begin{barticle}[auto:STB|2014/06/18|12:29:53]
\bauthor{\bsnm{Gwynn},~\bfnm{R.~C.}\binits{R.~C.}},
\bauthor{\bsnm{Burnett},~\bfnm{R.~T.}\binits{R.~T.}} \AND
\bauthor{\bsnm{Thurston},~\bfnm{G.~D.}\binits{G.~D.}}
(\byear{2000}).
\btitle{A time-series analysis of acidic particulate matter and daily
mortality and morbidity in the Buffalo, New York, region}.
\bjournal{Environ. Health Perspect.}
\bvolume{108}
\bpages{125--133}.
\end{barticle}
%
\bptok{imsref}%
\endbibitem

%b19 #&#
\bibitem[\protect\citeauthoryear{Horvitz and Thompson}{1952}]{horvitz1952}
%
\begin{barticle}[mr]
\bauthor{\bsnm{Horvitz},~\bfnm{D.~G.}\binits{D.~G.}} \AND
\bauthor{\bsnm{Thompson},~\bfnm{D.~J.}\binits{D.~J.}}
(\byear{1952}).
\btitle{A generalization of sampling without replacement from a finite
universe}.
\bjournal{J. Amer. Statist. Assoc.}
\bvolume{47}
\bpages{663--685}.
\bid{issn={0162-1459}, mr={0053460}}
\end{barticle}
%
\bptok{imsref}%
% NOT OUTPUTED:
% issn = 0162-1459
% fjournal = Journal of the American Statistical Association
\endbibitem

%b20 #&#
\bibitem[\protect\citeauthoryear{Kloog et~al.}{2012}]{kloog2012incorporating}
%
\begin{barticle}[auto:STB|2014/06/18|12:29:53]
\bauthor{\bsnm{Kloog},~\bfnm{Itai}\binits{I.}},
\bauthor{\bsnm{Nordio},~\bfnm{Francesco}\binits{F.}},
\bauthor{\bsnm{Coull},~\bfnm{Brent~A.}\binits{B.~A.}} \AND
\bauthor{\bsnm{Schwartz},~\bfnm{Joel}\binits{J.}}
(\byear{2012}).
\btitle{Incorporating local land use regression and satellite aerosol
optical depth in a hybrid model of spatiotemporal {PM}2.5 exposures in
the Mid-{A}tlantic states}.
\bjournal{Environmental Science \& Technology}
\bvolume{46}
\bpages{11913--11921}.
\end{barticle}
%
\bptok{imsref}%
% NOT OUTPUTED:
% number = 21
\endbibitem

%b21 #&#
\bibitem[\protect\citeauthoryear{Lawless, Kalbfleisch and
Wild}{1999}]{lawless1999semiparametric}
%
\begin{barticle}[mr]
\bauthor{\bsnm{Lawless},~\bfnm{J.~F.}\binits{J.~F.}},
\bauthor{\bsnm{Kalbfleisch},~\bfnm{J.~D.}\binits{J.~D.}} \AND
\bauthor{\bsnm{Wild},~\bfnm{C.~J.}\binits{C.~J.}}
(\byear{1999}).
\btitle{Semiparametric methods for response-selective and missing data
problems in regression}.
\bjournal{J. R. Stat. Soc. Ser. B Stat. Methodol.}
\bvolume{61}
\bpages{413--438}.
\bid{doi={10.1111/1467-9868.00185}, issn={1369-7412}, mr={1680310}}
\end{barticle}
%
\bptok{imsref}%
% NOT OUTPUTED:
% issn = 1369-7412
% url = http://dx.doi.org/10.1111/1467-9868.00185
% number = 2
% fjournal = Journal of the Royal Statistical Society. Series B.
%Statistical Methodology
\endbibitem

%b22 #&#
\bibitem[\protect\citeauthoryear{Le and Zidek}{2006}]{le2006statistical}
%
\begin{bbook}[mr]
\bauthor{\bsnm{Le},~\bfnm{Nhu~D.}\binits{N.~D.}} \AND
\bauthor{\bsnm{Zidek},~\bfnm{James~V.}\binits{J.~V.}}
(\byear{2006}).
\btitle{Statistical Analysis of Environmental Space--Time Processes}.
\bpublisher{Springer},
\blocation{New York}.
\bid{mr={2223933}}
\end{bbook}
%
\bptok{imsref}%
% NOT OUTPUTED:
% isbn = 0-387-26209-1; 978-0387-26209-3
% fpage = xvi+341
\endbibitem

%b23 #&#
\bibitem[\protect\citeauthoryear{Lee and Hirose}{2010}]{lee2010semi}
%
\begin{barticle}[mr]
\bauthor{\bsnm{Lee},~\bfnm{Alan}\binits{A.}} \AND
\bauthor{\bsnm{Hirose},~\bfnm{Yuichi}\binits{Y.}}
(\byear{2010}).
\btitle{Semi-parametric efficiency bounds for regression models under
response-selective sampling: The profile likelihood approach}.
\bjournal{Ann. Inst. Statist. Math.}
\bvolume{62}
\bpages{1023--1052}.
\bid{doi={10.1007/s10463-008-0205-1}, issn={0020-3157}, mr={2729153}}
\end{barticle}
%
\bptok{imsref}%
% NOT OUTPUTED:
% issn = 0020-3157
% url = http://dx.doi.org/10.1007/s10463-008-0205-1
% number = 6
% coden = AISXAD
% fjournal = Annals of the Institute of Statistical Mathematics
\endbibitem

%b24 #&#
\bibitem[\protect\citeauthoryear{Liang and
Zeger}{1986}]{liang1986longitudinal}
%
\begin{barticle}[mr]
\bauthor{\bsnm{Liang},~\bfnm{Kung~Yee}\binits{K.~Y.}} \AND
\bauthor{\bsnm{Zeger},~\bfnm{Scott~L.}\binits{S.~L.}}
(\byear{1986}).
\btitle{Longitudinal data analysis using generalized linear models}.
\bjournal{Biometrika}
\bvolume{73}
\bpages{13--22}.
\bid{doi={10.1093/biomet/73.1.13}, issn={0006-3444}, mr={0836430}}
\end{barticle}
%
\bptok{imsref}%
% NOT OUTPUTED:
% issn = 0006-3444
% url = http://dx.doi.org/10.1093/biomet/73.1.13
% number = 1
% coden = BIOKAX
% fjournal = Biometrika
\endbibitem

%b25 #&#
\bibitem[\protect\citeauthoryear{Monk and Munro}{2010}]{monk11}
%
\begin{bbook}[auto:STB|2014/06/18|12:29:53]
\bauthor{\bsnm{Monk},~\bfnm{P.}\binits{P.}} \AND
\bauthor{\bsnm{Munro},~\bfnm{L.~J.}\binits{L.~J.}}
(\byear{2010}).
\btitle{Maths for Chemistry: A Chemists Toolkit of Calculations},
\bedition{2nd} ed.
\bpublisher{Oxford Univ. Press},
\blocation{Oxford}.
\end{bbook}
%
\bptok{imsref}%
\endbibitem

%b26 #&#
\bibitem[\protect\citeauthoryear{Mueller}{1994}]{mueller1994characterization}
%
\begin{barticle}[auto:STB|2014/06/18|12:29:53]
\bauthor{\bsnm{Mueller},~\bfnm{Stephen~F.}\binits{S.~F.}}
(\byear{1994}).
\btitle{Characterization of ambient ozone levels in the Great Smoky
Mountains National Park}.
\bjournal{Journal of Applied Meteorology}
\bvolume{33}
\bpages{465--472}.
\end{barticle}
%
\bptok{imsref}%
% NOT OUTPUTED:
% number = 4
\endbibitem

%b27 #&#
\bibitem[\protect\citeauthoryear{Muir and Laxen}{1995}]{muir1995black}
%
\begin{barticle}[auto:STB|2014/06/18|12:29:53]
\bauthor{\bsnm{Muir},~\bfnm{D.}\binits{D.}} \AND
\bauthor{\bsnm{Laxen},~\bfnm{D.~P.~H.}\binits{D.~P.~H.}}
(\byear{1995}).
\btitle{Black smoke as a surrogate for PM$_{10}$ in health studies?}
\bjournal{Atmospheric Environment}
\bvolume{29}
\bpages{959--962}.
\end{barticle}
%
\bptok{imsref}%
% NOT OUTPUTED:
% number = 8
\endbibitem

%b28 #&#
\bibitem[\protect\citeauthoryear{Ott}{1990}]{ott1990}
%
\begin{barticle}[auto:STB|2014/06/18|12:29:53]
\bauthor{\bsnm{Ott},~\bfnm{W.}\binits{W.}}
(\byear{1990}).
\btitle{A physical explanation of the lognormality of pollutant
concentrations}.
\bjournal{Journal of the Air \& Waste Management Association}
\bvolume{40}
\bpages{1378--1383}.
\end{barticle}
%
\bptok{imsref}%
% NOT OUTPUTED:
% number = 10
\endbibitem

%b29 #&#
\bibitem[\protect\citeauthoryear{Pati, Reich and
Dunson}{2011}]{pati2011bayesian}
%
\begin{barticle}[mr]
\bauthor{\bsnm{Pati},~\bfnm{D.}\binits{D.}},
\bauthor{\bsnm{Reich},~\bfnm{B.~J.}\binits{B.~J.}} \AND
\bauthor{\bsnm{Dunson},~\bfnm{D.~B.}\binits{D.~B.}}
(\byear{2011}).
\btitle{Bayesian geostatistical modelling with informative sampling locations}.
\bjournal{Biometrika}
\bvolume{98}
\bpages{35--48}.
\bid{doi={10.1093/biomet/asq067}, issn={0006-3444}, mr={2804208}}
\end{barticle}
%
\bptok{imsref}%
% NOT OUTPUTED:
% issn = 0006-3444
% url = http://dx.doi.org/10.1093/biomet/asq067
% number = 1
% coden = BIOKAX
% fjournal = Biometrika
\endbibitem

%b30 #&#
\bibitem[\protect\citeauthoryear{Pfeffermann}{1993}]{pfeffermann93}
%
\begin{bmisc}[auto:STB|2014/06/18|12:29:53]
\bauthor{\bsnm{Pfeffermann},~\bfnm{D.}\binits{D.}}
(\byear{1993}).
\bhowpublished{The role of sampling weights when modeling survey data.
\emph{International Statistical Review/Revue Internationale de
Statistique} 317--337}.
\end{bmisc}
%
\bptok{imsref}%
% NOT OUTPUTED:
% sortkey = Pfeffermann(1993
% howpublished =
\endbibitem

%b31 #&#
\bibitem[\protect\citeauthoryear{Rao, Scott and Skinner}{1998}]{rao1998quasi}
%
\begin{barticle}[auto:STB|2014/06/18|12:29:53]
\bauthor{\bsnm{Rao},~\bfnm{J.~N.~K.}\binits{J.~N.~K.}},
\bauthor{\bsnm{Scott},~\bfnm{A.~J.}\binits{A.~J.}} \AND
\bauthor{\bsnm{Skinner},~\bfnm{C.~J.}\binits{C.~J.}}
(\byear{1998}).
\btitle{Quasi-score tests with survey data}.
\bjournal{Statist. Sinica}
\bvolume{8}
\bpages{1059--1070}.
\end{barticle}
%
\bptok{imsref}%
\endbibitem

%b32 #&#
\bibitem[\protect\citeauthoryear{Samet et~al.}{2000}]{samet00particulate}
%
\begin{barticle}[auto:STB|2014/06/18|12:29:53]
\bauthor{\bsnm{Samet},~\bfnm{J.~M.}\binits{J.~M.}},
\bauthor{\bsnm{Dominici},~\bfnm{F.}\binits{F.}},
\bauthor{\bsnm{Curriero},~\bfnm{F.~C.}\binits{F.~C.}},
\bauthor{\bsnm{Coursac},~\bfnm{I.}\binits{I.}} \AND
\bauthor{\bsnm{Zeger},~\bfnm{S.~L.}\binits{S.~L.}}
(\byear{2000}).
\btitle{Fine particulate air pollution and mortality in 20 {U.S.}
cities, 1987--1994}.
\bjournal{N. Engl. J. Med.}
\bvolume{343}
\bpages{1742--1749}.
\end{barticle}
%
\bptok{imsref}%
\endbibitem

%b33 #&#
\bibitem[\protect\citeauthoryear{S\"arndal, Swensson and
Wretman}{2003}]{sarndal03}
%
\begin{bbook}[mr]
\bauthor{\bsnm{S\"arndal},~\bfnm{Carl-Erik}\binits{C.-E.}},
\bauthor{\bsnm{Swensson},~\bfnm{Bengt}\binits{B.}} \AND
\bauthor{\bsnm{Wretman},~\bfnm{Jan}\binits{J.}}
(\byear{2003}).
\btitle{Model Assisted Survey Sampling}.
\bpublisher{Springer},
\blocation{New York}.
\bptnote{check year}%
\end{bbook}
%
\bptok{imsref}%
% NOT OUTPUTED:
% isbn = 0-387-97528-4
% url = http://dx.doi.org/10.1007/978-1-4612-4378-6
% fpage = xvi+694
\endbibitem

%b34 #&#
\bibitem[\protect\citeauthoryear{Schumacher and
Zidek}{1993}]{schumacher1993using}
%
\begin{barticle}[mr]
\bauthor{\bsnm{Schumacher},~\bfnm{Peter}\binits{P.}} \AND
\bauthor{\bsnm{Zidek},~\bfnm{James~V.}\binits{J.~V.}}
(\byear{1993}).
\btitle{Using prior information in designing intervention detection
experiments}.
\bjournal{Ann. Statist.}
\bvolume{21}
\bpages{447--463}.
\bid{doi={10.1214/aos/1176349036}, issn={0090-5364}, mr={1212187}}
\end{barticle}
%
\bptok{imsref}%
% NOT OUTPUTED:
% issn = 0090-5364
% url = http://dx.doi.org/10.1214/aos/1176349036
% number = 1
% coden = ASTSC7
% fjournal = The Annals of Statistics
\endbibitem

%b35 #&#
\bibitem[\protect\citeauthoryear{Scott and Wild}{2011}]{scott11}
%
\begin{barticle}[mr]
\bauthor{\bsnm{Scott},~\bfnm{Alastair~J.}\binits{A.~J.}} \AND
\bauthor{\bsnm{Wild},~\bfnm{Chris~J.}\binits{C.~J.}}
(\byear{2011}).
\btitle{Fitting regression models with response-biased samples}.
\bjournal{Canad. J. Statist.}
\bvolume{39}
\bpages{519--536}.
\bid{doi={10.1002/cjs.10114}, issn={0319-5724}, mr={2842429}}
\end{barticle}
%
\bptok{imsref}%
% NOT OUTPUTED:
% issn = 0319-5724
% url = http://dx.doi.org/10.1002/cjs.10114
% number = 3
% fjournal = The Canadian Journal of Statistics. La Revue Canadienne de
%Statistique
\endbibitem

%b36 #&#
\bibitem[\protect\citeauthoryear{Shaddick and Zidek}{2014}]{shaddick14}
%
\begin{barticle}[auto:STB|2014/06/18|12:29:53]
\bauthor{\bsnm{Shaddick},~\bfnm{G.}\binits{G.}} \AND
\bauthor{\bsnm{Zidek},~\bfnm{J.~V.}\binits{J.~V.}}
(\byear{2014}).
\btitle{A case study in preferential sampling: Long term
monitoring of air pollution.}
\bjournal{Spatial Statistics}
\bvolume{9}
\bpages{51--65}.
\end{barticle}
%
\bptok{imsref}%
% NOT OUTPUTED:
% sortkey = Shaddick
% howpublished =
\endbibitem

%b37 #&#
\bibitem[\protect\citeauthoryear{Stehman and
Overton}{1994}]{stehman1994comparison}
%
\begin{barticle}[auto:STB|2014/06/18|12:29:53]
\bauthor{\bsnm{Stehman},~\bfnm{Stephen~V.}\binits{S.~V.}} \AND
\bauthor{\bsnm{Overton},~\bfnm{W.~Scott}\binits{W.~S.}}
(\byear{1994}).
\btitle{Comparison of variance estimators of the Horvitz--{T}hompson
estimator for randomized variable probability systematic sampling}.
\bjournal{J. Amer. Statist. Assoc.}
\bvolume{89}
\bpages{30--43}.
\end{barticle}
%
\bptok{imsref}%
% NOT OUTPUTED:
% number = 425
\endbibitem

%b38 #&#
\bibitem[\protect\citeauthoryear{Verhoeff et al.}{1996}]{verhoeff96air}
%
\begin{barticle}[auto:STB|2014/06/18|12:29:53]
\bauthor{\bsnm{Verhoeff},~\bfnm{A.~P.}\binits{A.~P.}},
\bauthor{\bsnm{Hoek},~\bfnm{G.}\binits{G.}},
\bauthor{\bsnm{Schwartz},~\bfnm{J.~H.}\binits{J.~H.}}
\AND
\bauthor{\bsnm{van~Wijnen},~\bfnm{J.}\binits{J.}}
(\byear{1996}).
\btitle{Air pollution and daily mortality in Amsterdam}.
\bjournal{Epidemiology}
\bvolume{7}
\bpages{225--230}.
\end{barticle}
%
\bptok{imsref}%
\endbibitem

%b39 #&#
\bibitem[\protect\citeauthoryear{Zidek and
Shaddick}{2012}]{shaddick2012unbiasing}
%
\begin{bmisc}[auto:STB|2014/06/18|12:29:53]
\bauthor{\bsnm{Zidek},~\bfnm{J.~V.}\binits{J.~V.}} \AND
\bauthor{\bsnm{Shaddick},~\bfnm{G.}\binits{G.}}
(\byear{2012}).
\bhowpublished{Unbiasing estimates from preferentially sampled spatial data.
Technical Report 268.
Univ. British Columbia, Vancouver, BC}.
\end{bmisc}
%
\bptok{imsref}%
\endbibitem

%b40 #&#
\bibitem[\protect\citeauthoryear{Zidek, Sun and
Le}{2000}]{zidek2000designing}
%
\begin{barticle}[mr]
\bauthor{\bsnm{Zidek},~\bfnm{James~V.}\binits{J.~V.}},
\bauthor{\bsnm{Sun},~\bfnm{Weimin}\binits{W.}} \AND
\bauthor{\bsnm{Le},~\bfnm{Nhu~D.}\binits{N.~D.}}
(\byear{2000}).
\btitle{Designing and integrating composite networks for monitoring
multivariate {G}aussian pollution fields}.
\bjournal{J. R. Stat. Soc. Ser. C. Appl. Stat.}
\bvolume{49}
\bpages{63--79}.
\bid{doi={10.1111/1467-9876.00179}, issn={0035-9254}, mr={1817875}}
\end{barticle}
%
\bptok{imsref}%
% NOT OUTPUTED:
% issn = 0035-9254
% url = http://dx.doi.org/10.1111/1467-9876.00179
% number = 1
% fjournal = Journal of the Royal Statistical Society. Series C.
%Applied Statistics
\endbibitem
\end{thebibliography}
\end{document}